\documentclass[11pt, a4paper, logo, copyright]{googledeepmind}

\pdftrailerid{redacted}

\makeatletter
\renewcommand\bibentry[1]{\nocite{#1}{\frenchspacing\@nameuse{BR@r@#1\@extra@b@citeb}}}
\makeatother

\usepackage{kantlipsum, lipsum}
\usepackage{dsfont}
\usepackage{gdm-colors}
\usepackage[utf8]{inputenc}   
\usepackage{newunicodechar}   
\usepackage{wasysym,marvosym}
\usepackage{ulem}
\usepackage{hyperref}       
\usepackage{algorithmicx}
\usepackage{algpseudocode}
\usepackage{multirow}
\usepackage{geometry} 
\usepackage[rightcaption]{sidecap} 
\usepackage{wrapfig} 
\usepackage{tikz}
\usetikzlibrary{shapes.geometric, arrows.meta, positioning, backgrounds, fit}
\usepackage{amsmath}
\usepackage{algorithm}
\usepackage{algpseudocode}
\usepackage{listings}
\usepackage{enumitem} 
\usepackage{lipsum}
\usepackage[most]{tcolorbox}
\definecolor{thinkcolor}{RGB}{227,196,144}
\definecolor{observecolor}{RGB}{153,201,227}
\definecolor{explorecolor}{RGB}{178,217,200}

\definecolor{boxBorderBlue}{RGB}{92, 124, 179}   
\definecolor{titleBrown}{RGB}{139, 69, 19}       
\definecolor{titleBgGray}{RGB}{240, 240, 240}    
\definecolor{textOrange}{RGB}{198, 90, 40}       
\definecolor{textGreen}{RGB}{85, 139, 47}        

\newtcolorbox{evolvingbox}[1][]{
    breakable,             
    enhanced, 
    colback=white, 
    colframe=boxBorderBlue, 
    boxrule=1.5pt, 
    arc=4mm, 
    fontupper=\linespread{1.5}\selectfont, 
    parbox=false, 
    fontupper=\scriptsize,                       
    before upper={
        \setlength{\parindent}{0pt}
        \setlist[itemize]{topsep=0pt, partopsep=0pt, itemsep=1pt, parsep=0pt, leftmargin=5pt}%
        \setlist[enumerate]{topsep=0pt, partopsep=0pt, itemsep=1pt, parsep=0pt, leftmargin=5pt}%
    },
    title={Instruct}, 
    attach boxed title to top left={xshift=1.5cm, yshift=-\tcboxedtitleheight/4}, 
    boxed title style={
        colback=titleBgGray, 
        colframe=titleBrown, 
        boxrule=1.5pt,       
        arc=3mm,             
        top=1mm, bottom=1mm, left=3mm, right=3mm, 
    },
    coltitle=titleBrown, 
    fonttitle=\normalsize\bfseries, 
    #1
}

\newcounter{caseexample}[section]

\newcounter{promptexample}[section]

\usepackage{array, multirow, tabularx, booktabs, makecell}
\usepackage{booktabs} 
\usepackage{caption}  
\usepackage[export]{adjustbox}

\newcommand{\assignmentQuestionName}{Question} 

\usepackage{booktabs}
\usepackage{arydshln}
\usepackage{dashrule}
\usepackage{twemojis}

\usepackage[authoryear, sort&compress, round]{natbib}

\usepackage{bbding}
\usepackage[T1]{fontenc}    
\usepackage{url}            
\usepackage{booktabs}       
\usepackage{nicefrac}       
\usepackage{microtype}      
\usepackage{amsmath}
\usepackage{graphicx}
\usepackage{multicol}
\usepackage[nameinlink]{cleveref}
\usepackage{bbm}
\usepackage{multirow}
\usepackage{soul}
\usepackage{float}
\usepackage{wrapfig}
\usepackage{blindtext}
\usepackage{tablefootnote}
\usepackage{amsfonts}
\usepackage[flushleft]{threeparttable}
\usepackage{colortbl}
\usepackage{mathtools}
\usepackage{bm}
\usepackage{CJKutf8}
\usepackage{makecell}
\usepackage{caption}
\usepackage{capt-of}
\usepackage{array}
\usepackage{calc}      
\usepackage{caption}   
\usepackage{subcaption}  
\usepackage[bottom]{footmisc}
\usepackage{fontawesome}
\usepackage{tabularx}        
\usepackage{siunitx}

\usepackage[dvipsnames,svgnames]{xcolor} 
\usepackage[most]{tcolorbox}             

\definecolor{TagTime}{HTML}{0077BB}      
\definecolor{TagPlatform}{HTML}{33BBEE}    
\definecolor{TagType}{HTML}{009988}        
\definecolor{TagDuration}{HTML}{EE7733}    
\definecolor{TagCount}{HTML}{CC3311}      
\definecolor{TagItem}{HTML}{EE3377}        

\newcommand{\TimeTag}[1]{\textcolor{TagTime}{\textbf{[TIME: #1]}}}
\newcommand{\PlatformTag}[1]{\textcolor{TagPlatform}{\textbf{[Source: #1]}}}
\newcommand{\TypeTag}[1]{\textcolor{TagType}{\textbf{[TYPE: #1]}}}

\newcommand{\CountTag}[1]{\textcolor{TagCount}{\textbf{[COUNT: #1]}}}

\newcommand{\Items}[1]{\textcolor{TagItem}{\texttt{#1}}}

\newcommand{\Separator}{\quad\color{gray!40}\vrule width 1.5pt\quad}

\definecolor{RoleColor}{HTML}{CCFFCC}     
\definecolor{InputColor}{RGB}{52, 152, 219} 
\definecolor{MandatoryColor}{HTML}{FFCCCC} 
\definecolor{PresetColor}{HTML}{FFEECC}   
\definecolor{OutputColor}{HTML}{FFFFCC}   
\definecolor{HeaderColor}{HTML}{C6E2E9}   

\newcommand{\SectionTag}[2]{%
    \textcolor{black}{\fboxsep=3pt\fboxrule=0.5pt\colorbox{#1}{\textbf{#2}}}
}

\newcolumntype{C}{>{\centering\arraybackslash}X}
\newcolumntype{L}{>{\raggedright\arraybackslash}X}

\title{UserGPT Technical Report}
\author[1]{Yunyi Xuan}
\author[1]{Hao Yi}
\author[1]{Fengling Mao}
\author[1]{Daye Cai}
\author[1]{Leikun Liang}
\author[1$\dagger$]{Xingsheng He}
\author[1]{Jiangnan Xie}
\author[1]{Guoshuai Wang}
\author[1]{Yushan Han}
\author[1]{Wenwen Guo}
\author[1]{Xiaoxiao Xu}
\author[1]{Lin Qu}
\correspondingauthor={Xingsheng He}

\affil[1]{Alibaba Group}
\date{} 

\begin{abstract}
Personalization Understanding from vast digital traces is a fundamental challenge. Conventional user profiling relies heavily on discriminative models and manual feature engineering to produce discrete profile attributes, often resulting in fragmented, logically inconsistent profiles that fail to generalize across long-tail behaviors. Therefore, we challenge the conventional tag-based approach by a generative paradigm. In this framework, we task Large Language Models (LLMs) with distilling a user's long and noisy behavioral history into a dense, narrative summary that captures their nuanced evolution. While LLMs present a promising paradigm for unified user understanding, their foundational reasoning capabilities in this domain remain underexplored. Experimental results show that even state-of-the-art LLMs exhibit significant limitations in complex and implicit personalization.

To this end, we introduce \textit{UserGPT}, a principled framework for enhancing the reasoning abilities of LLMs in persona understanding and generating comprehensive profile attributes and summary. Acknowledging the scarcity of real-world behavior data, our framework begins with a novel \textit{User Behavior Simulation Engine} to generate realistic, complex user trajectories. These raw logs are then processed by our \textit{Data-Centric Semantization} module, which transforms heterogeneous behavioral data into structured, semantically coherent inputs, effectively mitigating inherent noise and sparsity. At the modeling layer, we introduce a \textit{Curriculum-Driven Post-Training} paradigm. This combines multi-stage Supervised Fine-Tuning (SFT) with our novel \textit{Dual-Filter Group Relative Policy Optimization (DF-GRPO)}, bolstering the model's ability to distill vast histories into insightful user profiles.

To evaluate these advanced reasoning capabilities, we construct the \textit{Holistic Persona Reasoning Bench (HPR-Bench)}, a new benchmark derived from our simulated data. On HPR-Bench, UserGPT achieves an Avg@10 score of $0.7325$ on tag prediction task and an Acc$_{\text{Ex}}$ score of $0.7528$ on summary generation task, a competitive performance compared with the state-of-the-art model, demonstrating a significant enhancement in its ability to reason about complex user profiles. The generated summary achieves up to $97.9\%$ data compression while preserving critical behavioral information. We claim that this establishes a solution that transcends the limitations of both traditional and naive LLM approaches, empowering next-generation AI applications and paving the way for personalized user-agent interactions.

\end{abstract}

\begin{document}
\begin{CJK*}{UTF8}{gkai}

\maketitle

\begin{figure}[!h]
    \centering

    \begin{subfigure}[t]{0.49\textwidth}
        \centering
        \includegraphics[width=\textwidth,height=0.23\textheight,keepaspectratio,valign=m]{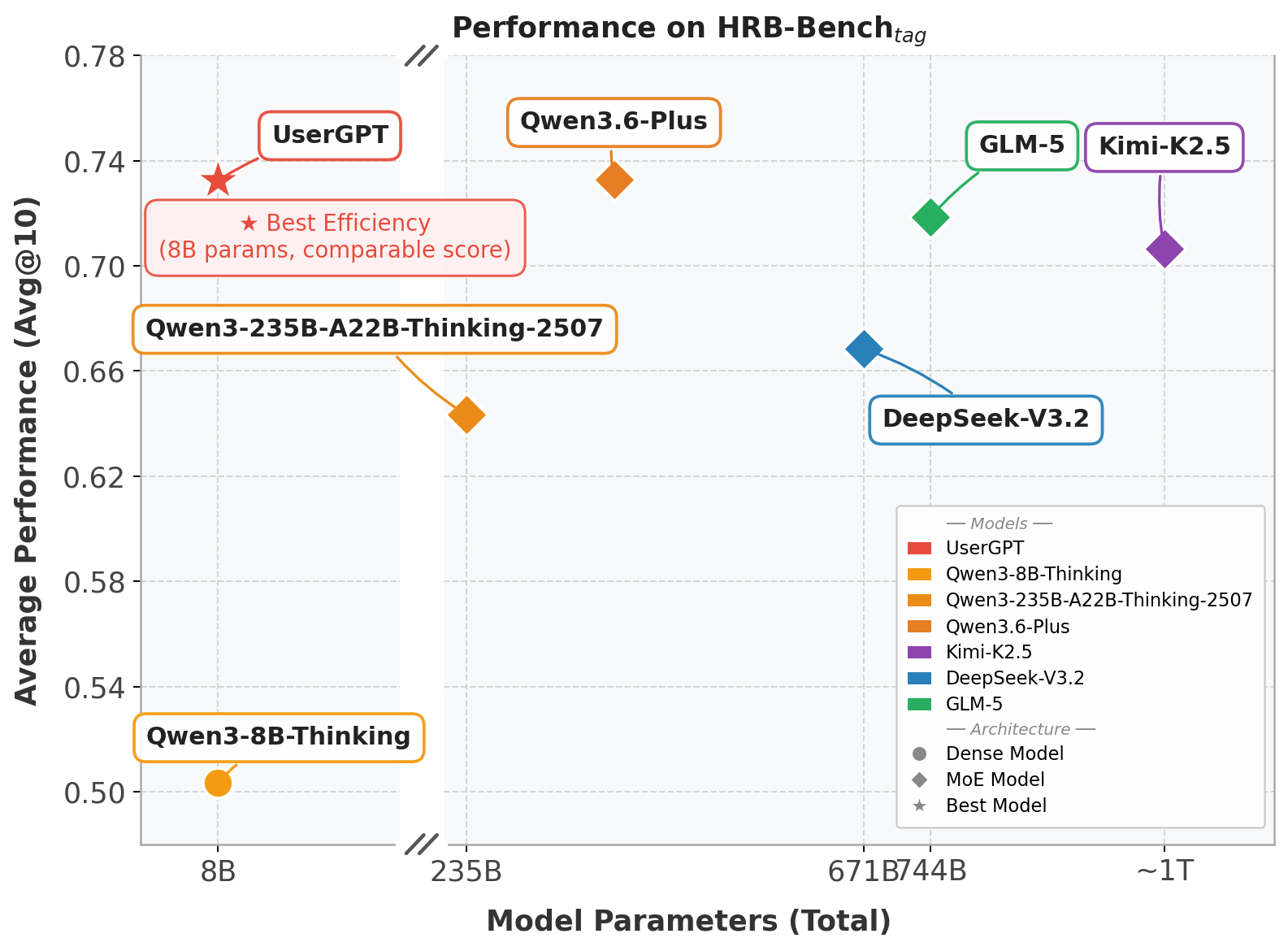}
        \label{fig:overall performance}
    \end{subfigure}%
    \hfill
    \begin{subfigure}[t]{0.49\textwidth}
        \centering
        \includegraphics[width=\textwidth,height=0.25\textheight,keepaspectratio,valign=m]{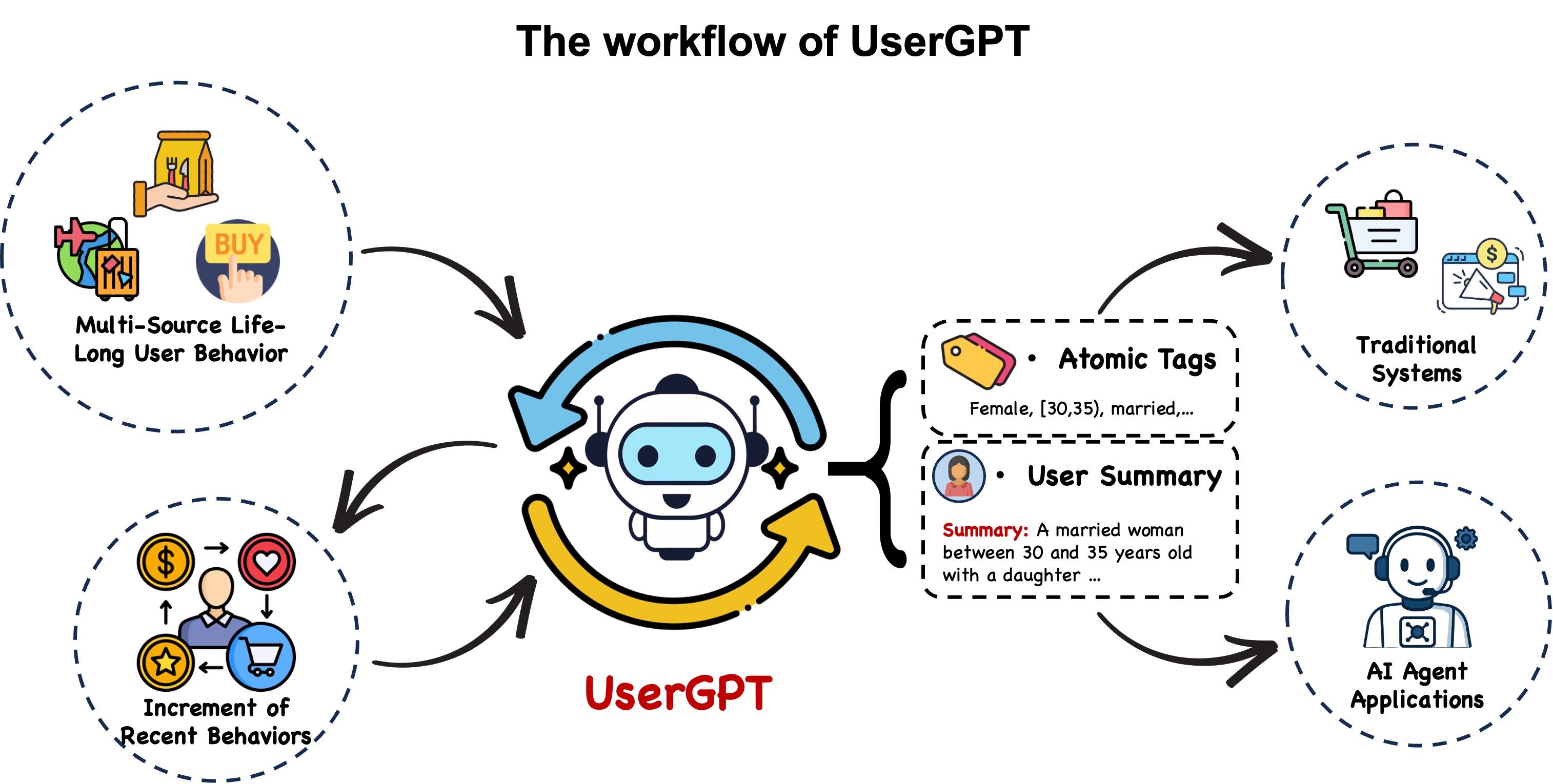}

        \label{fig:high-entropy}
    \end{subfigure}
    \vspace{-0.2cm}
    \caption{The overview of the proposed UserGPT for enhancing the personalization reasoning abilities.}
    \label{fig:entropy-comparison}
\end{figure}

\newpage
\setcounter{tocdepth}{2} 

\tableofcontents 

\newpage

\section{Introduction}
\label{sec:introduction}
With the rapid growth of services based on Large Language Models (LLMs), deep user understanding has emerged as the fundamental cornerstone for \textbf{personalized AI applications}. These range from personalized recommendation systems (\cite{sabouri2025towards, fabbri2025evaluating}) to intelligent chatbots and virtual assistants (\cite{team2025amap, wang2026productresearchtrainingecommercedeep}). As a critical component of user understanding, user profiling aims to distill structured features and insights from vast, noisy behavioral data. Over the past decades, this field has evolved from shallow, rule-based statistical methods (\cite{career_prediction_data_mining, gender_machine_learning, user_profiling_by_linear_regression}) toward discriminative prediction models powered by deep learning architectures (\cite{contrastive-multi-behavior, LURM}).  However, as digital ecosystems grow in complexity and user activities explode in volume, this prevailing discriminative paradigm is reaching a critical inflection point.

The core limitation of discriminative paradigm stems from its fundamental design, which is oriented towards producing a collection of discrete and isolated tags. This tag-centric approach inevitably forces a siloed modeling strategy where independent models are trained for each attribute, resulting in prohibitive drawbacks. Most notably, it frequently produces logically inconsistent user profiles, such as classifying a user as 'unmarried' while also 'having children'. Furthermore, it shows poor generalization to long-tail behaviors and fails to adapt to open-ended persona dimensions like evolving brand preferences.

The rise of LLMs, characterized by their vast world knowledge and sophisticated reasoning capabilities, offers a transformative paradigm shift from discrete classification towards ``generative paradigm'' to overcome these limitations. This approach treats the complex sequence of user behaviors as a unique form of ``language'' and directly generates user profiles through language modeling (\cite{Conf-Profile, e-commerce-user-profile}). Recent studies have investigated various strategies for constructing user profiles through prompting, fine-tuning (\cite{shi2025you, lu2025prompt}), and synthetic data generation (\cite{ge2024scaling, prottasha2025user}). 
Despite this progress, a naive application of LLMs to this task presents fundamental challenges. Even state-of-the-art LLMs still struggle  with the complex user understanding reasoning, exhibiting critical limitations in: 
\begin{itemize}
    \item \textbf{Data Scarcity and Usability Obstacles:} Enhancing the user understanding capabilities of LLMs fundamentally depends on training them with specifically designed data, yet the acquisition of such data presents a significant challenge. On one side, due to privacy and secrecy concerns, the authentic multi-year behavioral histories across multi-platforms remain inaccessible. On the other, these noisy and convoluted stream of logs are ill-suited for effective training.
    \item \textbf{Model-level Cognitive Obstacles:} Even frontier LLMs such as GPT-5, Gemini 3, or Qwen3.6-plus struggle with the complex cognitive reasoning required for accurate user profiling. User behavior histories are inherently noisy and filled with subtle cues that are difficult for models to access or interpret (\cite{liu2024lost, baker2024lost}). Additionally, LLMs often fail to recognize temporal dynamics, which prevents them from accurately tracking the evolution of a user's persona over time (\cite{xiong2024large, fatemi2024test}). These deficiencies lead to temporal reasoning errors, such as failing to infer a child's correct age from historical purchase data; dimensional conflicts, such as generating logically contradictory attribute combinations; and ambiguous behavior attribution, such as confusing gift-giving with personal consumption. As for practical industrial implementation, the ever-increasing user base and behavioral context window requirements pose a prohibitive engineering challenge to the deployment of LLMs in production environments.
\end{itemize}

To systematically tackle these challenges, we introduce UserGPT, a specialized LLM that begins with a robust data foundation before advancing to model training.

\textbf{Corpus Curation: User Behavior Simulation and Data-Centric Semantization.} We begin by tackling the data scarcity problem through the proposed User Behavior Simulation Engine, which crafts realistic, multi-year user life stories with controlled complexity and noise, providing the high-fidelity user behavior logs. With a source of realistic data secured, we then turn to the data usability challenge. Our Data-Centric Semantization pipeline acts as a crucial translator, transforming the raw, unintelligible logs into a structured, LLM-friendly narrative. This process primarily involves refining individual entities to be semantically coherent and weaving them into a cohesive behavioral corpus.

\textbf{UserGPT: Enhancing Cognitive Reasoning through Curriculum-Driven Post-Training.} 
To mitigate model-level cognitive challenges, our specialized training strategy for UserGPT recasts the user profiling task as a dual-objective problem: generating both structured tags and a narrative summary. Through a multi-stage process involving Supervised Fine-Tuning (SFT) and Reinforcement Learning (RL) (\cite{shao2024deepseekmath, ouyang2022training, schulman2017proximal}), guided by a multi-level curriculum (\cite{liu2024let, wang2024curriculum}), we systematically enhance UserGPT’s capability to perform temporal reasoning and resolve logical conflicts within extensive behavioral contexts. This enables the model to accurately track evolving personas. Building upon the reasoning capabilities of UserGPT, we introduce an ``Incremental Profiling'' paradigm to solve the computational update overhead, providing a cost-effective path for industrial user profiling.

Finally, to rigorously evaluate the capability of LLMs in capturing and synthesizing comprehensive user personas, we introduce the \textbf{Holistic Persona Reasoning Benchmark (HPR-Bench)}. This benchmark is specifically designed to assess models across two critical dimensions: atomic portrait tags (HPR-Bench\textsubscript{tag}) and user profiling summarization (HPR-Bench\textsubscript{sum}). The cornerstone of HPR-Bench lies in its unprecedented data quality and the rigorous multi-stage pipeline employed during its construction, which involves pre-annotation, difficulty stratification, automated quality control, and human verification. The resulting HPR-Bench represents one of the most reliable evaluation sets in the field of user modeling. Exhaustive experimental results demonstrate the superior performance of UserGPT in complex user understanding and persona reasoning compared to existing baselines.

In summary, this work presents a foundational exploration of generative lifelong user profiling.  We challenge the classical, tag-centric paradigm by reframing user profiling as a unified generative task: distilling vast, noisy behavioral logs into both structured tags and a narrative summary. The resulting narrative summaries can serve as a pluggable memory module for next-generation AI agents, paving the way for truly personalized user interactions. The main contributions of this paper are fourfold:

\begin{itemize}
    \item \textbf{An Elaborate‌ Data Curation Framework:} To overcome the critical challenges of data scarcity and usability, we first propose a User Behavior Simulation Engine to generate realistic, long-term user trajectories. We then introduce a Data-Centric Semantization pipeline to transform raw, noisy logs into a structured, LLM-friendly format, providing a solid foundation for this research.
    
    \item \textbf{An Advanced Training Methodology for Cognitive Enhancement:} We propose \textbf{UserGPT}, a specialized LLM optimized via a curriculum-driven, multi-stage training strategy. UserGPT significantly enhances the model's capability to perform temporal reasoning, resolve logical conflicts, and interpret the subtle, implicit behavioral cues.
    
    \item \textbf{Establishment of HPR-Bench:} We construct \textbf{HPR-Bench}, a comprehensive evaluation benchmark designed to bridge the scarcity of metrics for complex user profiling. It provides the research community with a standardized, robust toolkit to measure and advance the generative understanding of user attributes over long time horizons.
    
    \item \textbf{Empirical Validation:} We demonstrate through exhaustive experiments that UserGPT significantly outperforms baseline methods in both efficiency and quality, particularly in complex cognitive reasoning tasks.
\end{itemize}

\section{Data Simulation and Data-Centric Semantization}
\label{sec:data}
A robust data foundation for user profiling must overcome two fundamental challenges: data scarcity and data usability. The acquisition of granular, long-term real-world behavioral data is nearly impossible. Simultaneously, the symbolic, event-driven nature and extreme length of raw interaction logs create a significant modality mismatch, rendering them semantically opaque to LLMs.

To address these challenges, we introduce a comprehensive, two-stage data preparation strategy. First, we utilize a sophisticated simulation framework to generate realistic, multi-year user traces, resolving the scarcity issue (Section \ref{sec:behavior simulation}). Second, we implement a Data-Centric Semantization pipeline that transforms these raw logs into LLM-friendly representations via micro-level \textit{Entity Refinement} (Section \ref{subsec:entity refinement}) and macro-level \textit{Behavioral Corpus Construction} (Section \ref{subsec:Behavioral Corpus Construction}).

\subsection{User Behavior Simulation}
\label{sec:behavior simulation}
To generate high-fidelity synthetic data, we develop a simulation framework consisting of four core modules: a Persona-Driven Agent, an Environment \& Interaction Agent, a Simulation Engine with Persona Evolution, and a Quality Assurance Mechanism.

\paragraph{Persona-Driven Agent}
Inspired by Belief-Desire-Intention Engine (\cite{lifesim}), we design a three-tiered Persona-Need-Intent user cognition model specifically tailored for consumer scenarios:
\begin{itemize}
    \item \textbf{Core Persona}: By integrating the 90-dimensional preference space from AlignX (\cite{AlignX}) and the 15-dimensional annotation system from SocioVerse (\cite{socioverse}), we construct a high-dimensional user profile encompassing demographics, Big Five personality traits, and psychological consumption needs.
    \item \textbf{Latent Needs}: Initialized from the "Desire Pool" of LifeSim(\cite{lifesim}), we extensively customize it for e-commerce scenarios. By leveraging LLMs to mine consumer intents from real user reviews, we created a specialized ``Consumption Desire Library''. The agent activates matching latent needs from this library based on the user's Core Persona and current life stage.
    \item \textbf{Specific Intent}: When triggered by environmental status (e.g., the shopping festival or social hotspots), these latent needs are crystallized into concrete intents with specific target platforms, action types, and expected outcomes.
\end{itemize}

The Persona-Driven Agent randomly selects a plausible triplet of (Core Persona, Latent Need, Specific Intent), completes any missing dimensions while ensuring the persona's internal consistency. It then generates the user's life trajectory by chronologically advancing through life stages (e.g., starting school, graduating, entering the workforce). Notably, to ensure data authenticity, we adjust the distribution of the final user pool to align with China's demographic statistics.

\paragraph{Environment \& Interaction Agent}
This agent is responsible for constructing a dynamic external world and serving as the interface for interaction.
\begin{itemize}
    \item \textbf{Dynamic Environment Simulation}: It simulates a dynamic spatio-temporal: temporal dimension, which advances daily and marks key events (e.g., 618, Double 11, back-to-school season, Spring Festival, social hotspots); and spatial dimension, which generates key POIs based on the user's life trajectory.
    \item  \textbf{Cross-Platform Tool Graph Interaction}: We adapt and extend concepts from VitaBench (\cite{vitabench}) to construct a Cross-Platform Tool Graph, which is applied to multi-platform interactions, covering e-commerce, food delivery, Online Travel Agency (OTA), and POI visits\footnote{For the POI platform, we utilize timestamps and corresponding POI types from the Foursquare dataset (\cite{Foursquare}).}. This significantly enhances the complexity and realism of the traces. 
\end{itemize}

\paragraph{Simulation Engine with Persona Evolution}
Simulation Engine employs a Markov Decision Process to model user behavior state transitions (e.g., idle → browsing → searching → ordering). At each timestep, it adjusts state transition probabilities based on the user's cognitive state and environmental status, ensuring an orderly and plausible simulation flow.

To model the dynamic evolution of user personas, we design the Persona Evolution Engine, which is maintained as one of the core capabilities of the Persona-Driven Agent:
\begin{itemize}
    \item \textbf{Event-Driven}: Triggered by significant life events detected by the system (e.g., pregnancy, relocation, career change), resulting in significant and long-term qualitative shifts in the Core Persona.
    \item \textbf{Behavior-Driven}: Executed at fixed intervals (e.g., monthly), fine-tuning dynamic preferences based on feedback from recent behaviors.
\end{itemize}

\paragraph{Quality Assurance Mechanism}
To ensure the authenticity of the synthetic user trajectories, we implement a quality assurance framework:
\begin{itemize}
    \item \textbf{Logical Consistency Validation}: Built-in automated rules to detect and rectify logical inconsistencies between behaviors, personas, platforms, and intents in real-time.
    \item \textbf{Authenticity via Noise Injection}: Simulating real-world uncertainties by injecting noise, such as operational errors (e.g., misclicks) and shared account scenarios.
    \item \textbf{Manual Sample Validation}: Periodic manual validation of sampled traces. These samples are assessed across multiple dimensions, such as behavioral plausibility and persona consistency. The findings are used to iteratively refine the generation models and validation rules.
\end{itemize}

\subsection{Entity Refinement: Micro-level Denoising} 
The raw logs generated by the simulation consist of "entities" (e.g., e-commerce products, travel hotels) that are inherently heterogeneous, originating from diverse platforms with distinct attributes and data structures, as shown in Table \ref{tab:refinement_examples}. Simply concatenating this raw, multi-source data creates a chaotic input stream that hinders model comprehension. To make these entities interpretable for LLMs, we address two primary obstacles:

\definecolor{atrri_token}{RGB}{52, 152, 219} 
\newcommand{\attri}[1]{\colorbox{atrri_token!30}{\strut #1}}
\newcommand{\core}[1]{\colorbox{yellow!30}{\strut #1}}

\begin{table}[!t]
\centering
\caption{Examples of Semantic Refinement with Semantic Tagging:\attri{descriptive modifier}, \core{core noun phrase}}
\label{tab:refinement_examples}
\small
\begin{tabularx}{\linewidth}{
    >{\hsize=0.20\hsize\raggedright\arraybackslash}X
    >{\hsize=0.15\hsize\raggedright\arraybackslash}X
    >{\hsize=0.33\hsize\raggedright\arraybackslash}X
    >{\hsize=0.33\hsize\raggedright\arraybackslash}X
}
\toprule
\textbf{Source} &\textbf{Problem} & \textbf{Raw Title} & \textbf{Refined Title} \\
\midrule
E-commerce &Semantic noise & 2023 autumn \& winter new slim fit mid-sleeve knit sweater for women, pullover sweater, base layer, gentle style, short style & \attri{Women's} \attri{slim-fit} \attri{short} \core{knit top} \\ \midrule
E-commerce &Semantic noise &  Bright Dairy full-fat high-calcium milk powder for adults, women and students – high in calcium and iron, nutritious breakfast milk for the whole family, 400g × 3 bags&  \attri{Bright Dairy} \attri{high-calcium} \attri{high-iron} \core{milk powder} \\ \midrule
OTA &Information sparsity & Ocean View Luxury Suite & \attri{Hailing Island, Guangdong}\allowbreak\attri{Province}-\attri{Budget Hotel}-\attri{Ocean View}-\core{Luxury Suite} \\ \midrule
Delivery &Information sparsity\& Semantic noise & Meng Meilong (Honeydew Melon Oolong) &\attri{Drinks}-\attri{SCHAGEE}-\core{Honeydew Melon Oolong}\\
\bottomrule
\end{tabularx}
\end{table}

\textbf{(1) Semantic Noise}, arises from marketing-driven language, particularly on e-commerce platforms where vendors embed promotional keywords into product titles. These ``soft labels'', such as ``pregnant women'' in ``wet wipes for pregnant women'' or ``student'' in ``student desk lamp'', misrepresent universally applicable products. For a model, these weak signals can cause attribution bias, leading it to incorrectly infer user attributes (e.g., assuming any buyer is pregnant). Moreover, this promotional clutter obscures the product's core identity and wastes computational resources on parsing irrelevant information.

\textbf{(2) Information Sparsity}, conversely, is prevalent on platforms with constantly emerging content, such as travel. New entities, such as a viral tourist spot emerging with only a name, are effectively unknown to an LLM as they post-date its pre-training knowledge cutoff. This creates a cold-start problem, as the model lacks the contextual metadata (e.g., location, type of scenery, available activities) needed for understanding. For instance, if a model encounters a trendy new spot named ``Anji's Little Iceland'' without any context, it cannot infer a user's underlying preference for nature-themed travel or minimalist landscapes. Such context-deficient entities remain isolated tokens, severely limiting the model's ability to capture and generalize user interests.

\begin{figure}[!t]
    \centering
    \includegraphics[width=0.95\textwidth]{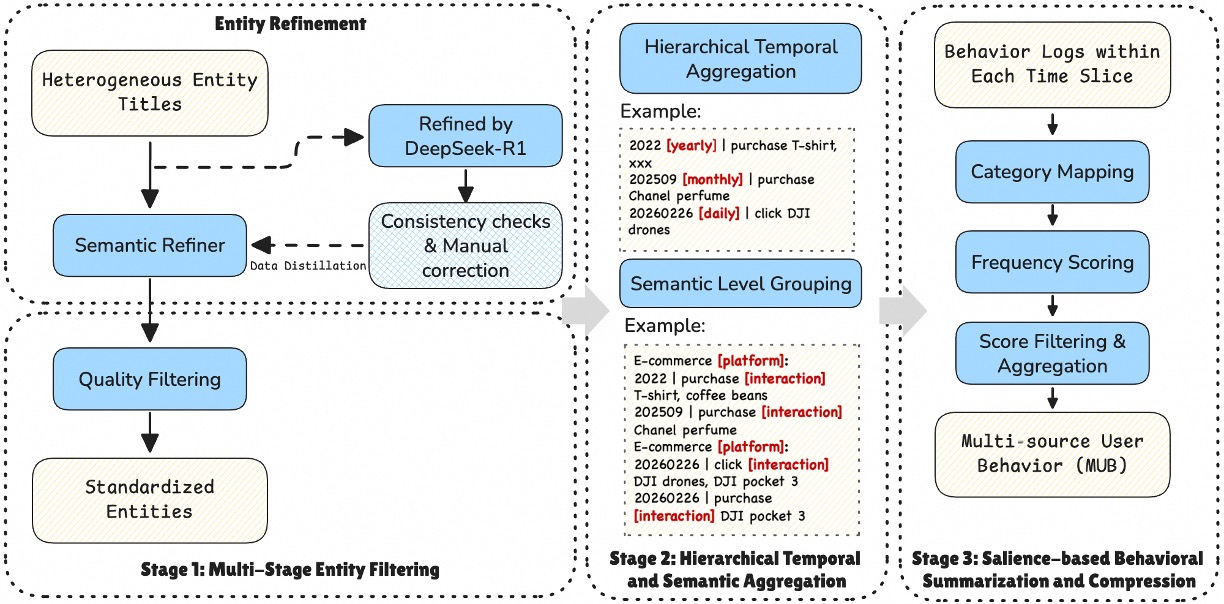}
    \caption{The entire pipeline of Data-Centric Semantization, including \textbf{Entity Refinement} mentioned in Section \ref{subsec:entity refinement} and \textbf{Behavior Corpus Construction} mentioned in Section \ref{subsec:Behavioral Corpus Construction}.}
    \label{fig: Data-Centric Semantization pipeline}
\end{figure}

 \paragraph{Methods.} To address the aforementioned challenges of semantic noise and information sparsity, we propose a unified Semantic Refiner built upon a fine-tuned, lightweight LLM (Qwen3-1.7B). The complete process is depicted in Figure \ref{fig: Data-Centric Semantization pipeline}. This refiner efficiently processes the large scale of entities by transforming heterogeneous inputs into a standardized, LLM-friendly representation. It employs a dual-strategy approach:

\label{subsec:entity refinement}
\subparagraph{Entity Rewriting (Denoising and Compression).} For noisy entities from sources like e-commerce, the refiner applies a ``rewrite-and-extract'' paradigm. It distills marketing-heavy titles into a canonical format, one core entity and up to three salient attributes, by stripping promotional noise while preserving core semantics (e.g., brand, audience).
\vspace{-0.5cm}
\subparagraph{Entity Enrichment (Normalization and Augmentation).} For sparse entities (e.g., from OTA), the refiner queries internal knowledge bases to augment them with critical metadata (e.g., categories, address) before compressing them into the same standardized format. This ensures all entities are structurally uniform and semantically comprehensible.
\vspace{-0.5cm}
\subparagraph{Training and Evaluation.} We synthesized training data using DeepSeek-R1, prompting it to generate refined outputs according to our schema, followed by consistency checks and manual correction. For evaluation, we constructed a high-quality test set of approximately 6,000 samples by expert annotation. On this benchmark, our lightweight refiner significantly outperformed both the large-scale DeepSeek-R1 and a traditional NER (Named Entity Recognition) with handcrafted rules baseline on BLEU and ROUGE scores (Table~\ref{tab:rewrite_eval}).

This refinement process (shown in Table~\ref{tab:refinement_examples}) reduces average token length by over 50\%, lowering the computational burden on downstream models. By focusing on core semantics, it also enhances the precision of user attribute prediction.

\begin{table}[htbp]
\centering
\caption{Evaluation Results of Semantic Refinement Models}
\label{tab:rewrite_eval}
\begin{tabular}{lcccc}
\toprule
\textbf{Method} & \multicolumn{2}{c}{\textbf{BLEU}} & \multicolumn{2}{c}{\textbf{ROUGE}} \\
\cmidrule(lr){2-3} \cmidrule(lr){4-5}
& \textbf{n-gram=1} & \textbf{n-gram=2} & \textbf{ROUGE-1} & \textbf{ROUGE-L} \\
\midrule
NER+Handcrafted Rules        & 0.4196 & 0.1870 & 0.4775 & 0.4631 \\
DeepSeek-R1    & 0.6213 & 0.3962 & 0.6691 & 0.6547 \\
Semantic Refiner (Ours)           & \textbf{0.6451} & \textbf{0.4294} & \textbf{0.6975} & \textbf{0.6787} \\
\bottomrule
\end{tabular}
\end{table}

\subsection{Behavioral Corpus Construction: Macro-level Structuring}
\label{subsec:Behavioral Corpus Construction}
While Entity Refinement cleans individual entities, we still face the challenge of prohibitive sequence length and unintelligible nature of user logs. We propose a novel \textbf{Hierarchical Spatio-Temporal Behavior Corpus Construction Paradigm}. This paradigm is designed to transform raw, noisy behavioral logs into structured, information-dense sequences that are amenable to LLMs. As illustrated in Figure \ref{fig: Data-Centric Semantization pipeline}, the entire preprocessing pipeline consists of three core stages:

\paragraph{Stage 1: Multi-Stage Entity Filtering.} Although the raw entities have been transformed into a standardized format via the process in Section \ref{subsec:entity refinement}, many may still lack practical significance. We therefore employ a multi-stage entity filtering mechanism to refine the candidate set. By combining NER models, category constraints, text complexity analysis, and stop-word lists, this stage systematically identifies and discards non-informative entities, ensuring that only high-quality, task-relevant entities are retained.

\paragraph{Stage 2: Hierarchical Temporal and Semantic Aggregation.} This stage organizes the user's behavioral stream in a manner consistent with human memory and cognition, operating along two dimensions:
\begin{itemize}
    \item \textbf{Semantic Level Grouping:} We group behaviors based on their source and interaction type (e.g., search, click, purchase, visit). This step aims to disentangle user intents across different life facets. For example, a ``purchase'' on an e-commerce service signifies strong user demand, whereas a ``travel'' on an OTA platform reveals the user's geographical preference.
    \item \textbf{Hierarchical Temporal Aggregation:} Inspired by human memory, we employ a coarse-to-fine temporal aggregation strategy. For long-term history, we aggregate only high-impact actions (e.g., purchases) at a coarse monthly or yearly granularity to capture stable life stages. Conversely, for short-term history, we retain a comprehensive set of behaviors, including searches, clicks, and purchases, at a fine-grained daily level to capture a high-fidelity view of emerging user intents. This dual-granularity approach allows the model to understand both stable, long-term trends and recent, dynamic shifts in user interest.
    
\end{itemize}
\paragraph{Stage 3: Salience-based Behavioral Summarization and Compression.} After initial sequential assembly, we introduce a behavior compression mechanism to adhere to the model's context window limitations and enhance the signal-to-noise ratio. This mechanism effectively functions as an implicit attention mechanism, guiding the model to focus on the most critical information within each period.
\begin{itemize}
    \item Specifically, within each time slice (e.g., a single day), all behaviors are mapped to a unified, predefined category taxonomy (e.g., Parenting \& Baby, Consumer Electronics, Outdoor Sports).
    \item We then compute the frequency of user interactions within each category, treating this frequency as a salience score.
    \item Finally, we retain only the core behaviors from the Top-K most salient categories, while filtering out anomalous patterns, such as synthetic noise behaviors intentionally injected during simulation and erroneous infinite loop sequences. This produces a periodic behavioral summary that is both compact and highly focused.

\end{itemize}
Ultimately, the user's long-term behavior is organized into a structured sequence, formatted as follows, which clearly presents the time, source, behavior type, frequency, and specific content:
\definecolor{box_bg}{RGB}{245, 250, 255} 
\definecolor{box_frame}{RGB}{52, 152, 219} 
\begin{tcolorbox}[
    enhanced,
    colback=box_bg, 
    colframe=box_frame, 
    fonttitle=\bfseries,
    title=Example of a Multi-source User Behavior (MUB),
    attach boxed title to top center={yshift=-2mm},
    boxed title style={
        colback=box_frame, 
        colframe=box_frame,
        arc=1mm,
    },
    arc=2mm, 
    boxrule=1pt, 
]
\PlatformTag{E-commerce} \quad
\TimeTag{2026-01-28} \quad
\TypeTag{Click} \quad
\CountTag{5}
\quad $|$ \quad 
\Items{Wet Wipes, Milk Powder, \dots}

\Separator

\PlatformTag{Delivery} \quad
\TimeTag{2025-12} \quad
\TypeTag{Purchase} \quad
\quad $|$ \quad
\Items{Starbucks Latte, \dots}

\Separator

\PlatformTag{Source$_n$} \quad
\TimeTag{Time$_n$} \quad
\TypeTag{Behavior$_n$} \quad
\CountTag{Frequency$_n$ (Optional)} \quad
\quad $|$ \quad
\Items{Item$_1$, Item$_2$ \dots}
\end{tcolorbox}

Through the proposed Data-Centric Semantization pipeline, raw behavioral logs can be successfully transformed into a high-density, structured format we term the Multi-source User Behavior (MUB), yielding transformative improvements in both efficiency and quality. Regarding quality, the Entity Refinement and Behavioral Corpus Construction eliminate non-informative noise, significantly enhancing the semantic readability and coherence of the behavioral data. On the efficiency front, this pipeline achieves a data reduction rate of over 75\% through entity denoising and salience-based compression. This drastically cuts the computational cost of long-context inference and makes it feasible to process extended user behavioral patterns within a finite context window in a production environment.

\section{Model Training \& Adaptation}
\label{sec:training}
\subsection{Problem Definition}
\label{subsec:problem_definition}
The primary objective of \textit{UserGPT} is to bridge the gap between legacy heuristic profiling and the growing need for explainable, consistent, and comprehensive user understanding. Conventional systems, which typically rely on shallow, rule-based statistical methods (e.g., classifying a user as ``non-single'' based on the purchase of couple-themed items), suffer from severe limitations:
\begin{itemize}
    \item \textbf{Compromised Label Quality}: Legacy systems frequently suffer from significant noise and logical conflicts. Rule-based methods lack the semantic depth to differentiate diverse behavioral intents, often misinterpreting accidental clicks or exploratory browsing as stable preferences (e.g., an unmarried user clicking on baby products being tagged as a parent). Furthermore, because profile attributes are generated using independent and fragmented rule sets, logical conflicts are common. For instance, a user being simultaneously tagged as a ``42-year-old'' and a ``college student'' when in reality they are a parent purchasing items for their child.
    \item \textbf{Limited Label Coverage}: The rigidity of rule-based systems prevents them from encompassing the full spectrum of user behaviors. Even when users exhibit high levels of activity, the system may fail to generate a meaningful profile if the specific behavior patterns do not strictly align with the predefined rules. Furthermore, explicit rules are fundamentally ill-equipped to model latent attributes, such as evolving user interest preferences, which often require a more sophisticated semantic understanding rather than rule-based logic.
    \item \textbf{Inefficient Production and Application}: In large-scale production, atomic labels are typically generated by passing the same raw input through a multitude of independent models. This ``one-model-per-label'' architecture leads to significant maintenance overhead and resource waste. Moreover, these black-box models offer poor explainability, making it difficult to trace the rationale behind specific tags. At the application level, different business units often use proprietary rules to ``stitch'' these atomic labels into a composite profile summary. This approach fails to capture the synergetic semantics between labels. 
\end{itemize}

Formally, let $U$ denote the MUB of a given user, encompassing multi-source interactions such as e-commerce transactions, video viewing logs, and other digital footprints, as detailed in Section~\ref{subsec:Behavioral Corpus Construction}. We define a set of $K$ atomic user profile attributes denoted as $\{y^a_1, \dots, y^a_K\}$. For each atomic attribute $y^a_k$, the corresponding value space is defined based on its data type:

\begin{itemize}
    \item \textbf{Categorical Attributes}: For discrete attributes, such as gender, the value space is defined as $\mathcal{Y}^a_k=\{y^a_{k,1},\dots,y^a_{k,m_k}\}\cup \{\text{NA}\}$, where $m_k$ denotes the number of possible categorical values and \text{NA} denotes an unknown value.
    \item \textbf{Open-domain Attributes}: For generative attributes like interest preferences, the value is drawn from an open-text space, denoted by $\mathcal{Y}^a_k = \mathcal{T}$, where $\mathcal{T}$ represents the space of natural language texts.
\end{itemize}

Beyond atomic attributes, the model generates a summary of the composite profile, denoted by $y^s$, which provides a holistic synthesis of the user’s long-term lifecycle characteristics.

The user profiling task is thus formulated as a joint multi-output classification and text generation problem. Let $\mathcal{Y}^a = \mathcal{Y}^a_1 \times \cdots \times \mathcal{Y}^a_K$ denote the joint output space of all atomic attributes, and let $\mathcal{Y}^s$ denote the composite profile summary space. We seek to learn a mapping function $f: U \rightarrow \mathcal{Y}^a \times \mathcal{Y}^s$. Given user data $U$, the function produces the predicted atomic attributes $\hat{\mathbf{y}}^a = (\hat{y}^a_1, \dots, \hat{y}^a_K)$ and a generated composite summary $\hat{y}^s$:
\begin{equation}
    (\hat{\mathbf{y}}^a, \hat{y}^s) = f(U),
\end{equation}
where $\hat{\mathbf{y}}^a \in \mathcal{Y}^a$ and $\hat{y}^s \in \mathcal{Y}^s$. The function $f(\cdot)$ is optimized to maintain logical consistency across generated atomic attributes and ensure factual alignment with the ground-truth user behaviors.

\subsection{Synthetic Data Quality Challenges}
\label{subsec:Data Quality Challenges}
Despite the abundance of raw behavioral data, the acquisition of high-quality annotations paired with explicit reasoning traces remains a bottleneck. Leveraging LLMs for synthetic data generation introduces four primary challenges:

\begin{enumerate}
    \item \textbf{Semantic Hallucination}: LLMs frequently produce factually incorrect inferences in specialized fields that require deep domain knowledge. For instance, when inferring a baby's precise age based on sequential stages of milk powder consumption, LLMs may fail to capture the developmental logic. Similarly, LLMs may misinterpret consumer intent by relying on superficial product associations rather than nuanced logic. For instance, when an adult with sensitive skin purchases baby wipes due to their stricter safety standards, LLMs might erroneously classify the consumer as a ``new parent''.
    \item \textbf{Contextual Fragmentation}: While standard LLMs demonstrate proficiency in isolated attribute inference, they often struggle to maintain logical consistency when processing extended behavioral sequences. This fragmentation manifests in three primary ways:
        \begin{itemize}
            \item \textit{Persona Contamination}: Failure to disentangle multi-user signals in shared account scenarios (e.g., a parent and child using the same device). This leads to the synthesis of contradictory personas, such as labeling a user as a ``42-year-old college student''.
            \item \textit{Temporal Evolution Blindness}: Models often treat behavioral snapshots in isolation, failing to track the natural progression of a user's life stage. For instance, LLMs might generate a profile that simultaneously attributes ``single'' and ``parenting'' status to a user because it cannot reconcile the transition from singlehood to parenthood over time.
            \item \textit{Intent Misattribution}: There is a significant challenge in distinguishing core personal preferences from external motivations. LLMs often misinterpret commercial behaviors (e.g., bulk-buying toys for resale) or altruistic actions (e.g., one-time gifting) as permanent personal traits, resulting in a fragmented and contradictory user profile.
        \end{itemize}
    \item \textbf{Intrinsic Data Noise}: LLMs often exhibit an ``over-interpretation'' bias, failing to distinguish between preference-driven actions and accidental behaviors. For instance, a single user might browse couple-themed items out of fleeting curiosity or accidentally click on an infant-related advertisement. Because LLMs lack the ability to weigh the ``significance'' of individual interactions, they tend to treat these transient signals as stable user traits, leading to the generation of unreliable profiles.
    \item \textbf{Trade-offs between Cost and Performance}: The synthesis of high-fidelity data involves a critical balance between computational cost and model performance. In the absence of ground-truth labels, achieving high-fidelity data typically necessitates resource-intensive processes such as multi-round consistency verification. Consequently, employing larger LLMs to produce high-fidelity data significantly escalates the overall computational cost.
\end{enumerate}

To address these challenges, we propose a curriculum-driven post-training paradigm that integrates optimized prompt engineering, dual quality verification, and difficulty-aware sampling.

\subsection{Curriculum-Driven Post-Training}
\label{subsec:Post-Training}
\begin{figure}[htbp]
    \centering 
    \includegraphics[width=0.95\textwidth]{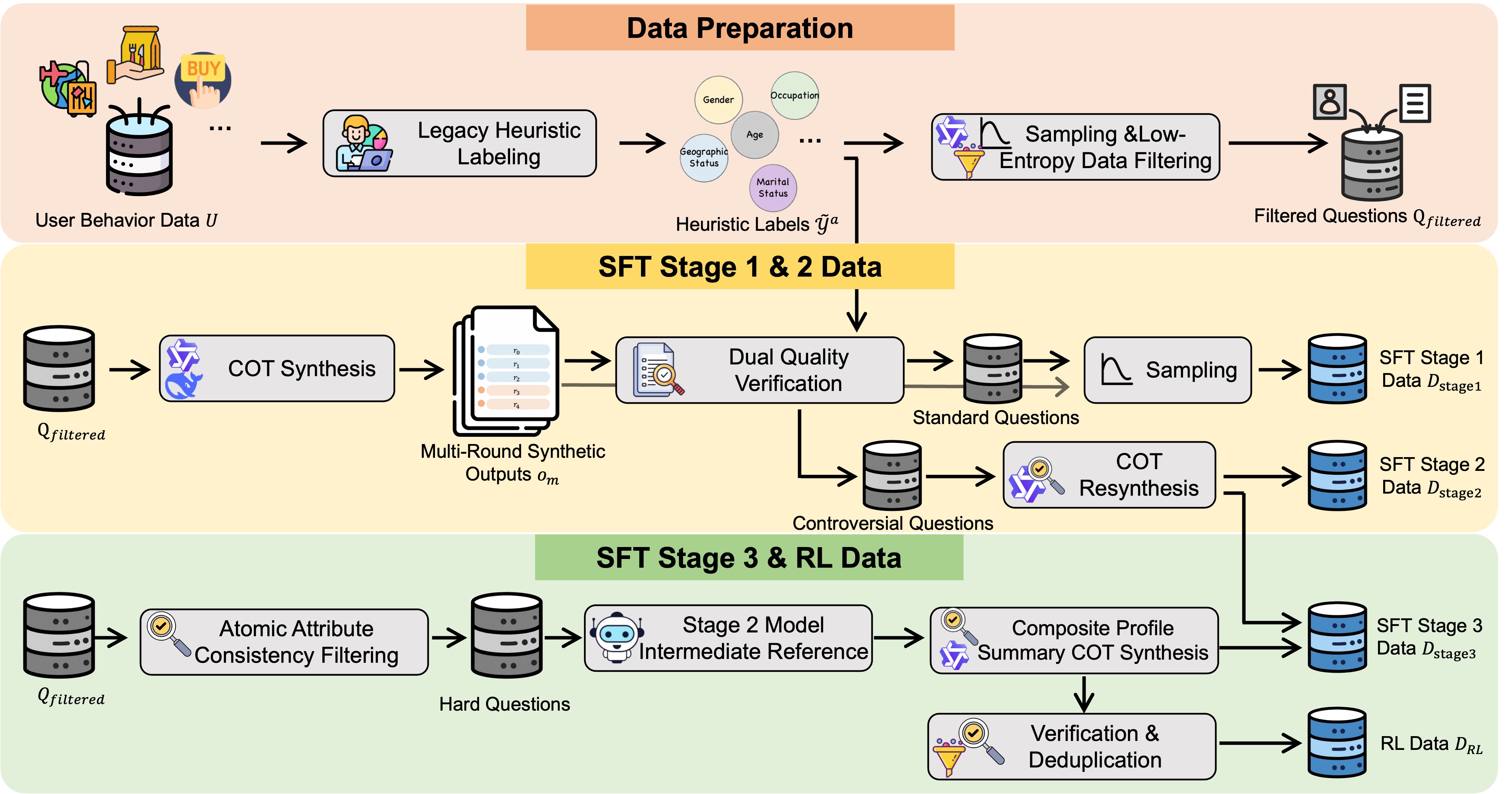} 
    \caption{The Data Engineering Pipeline for Curriculum-Driven Post-Training. The process starts with a \textbf{Data Preparation} phase to establish a reliable set of pseudo-labels and a balanced user pool. \textbf{SFT Stages 1 and 2} are designed to build the model's foundational atomic attribute inference skills through synthesized CoT samples. \textbf{SFT Stage 3} and \textbf{RL} are employed to develop the model's proficiency in generating composite profile summaries.} 
    \label{fig:data_process} 
\end{figure}

The post-training pipeline, as illustrated in Figure \ref{fig:data_process}, follows a structured progression from basic alignment to complex reasoning. This curriculum-driven data synthesis pipeline enables a transition from trustworthy data to pedagogically effective data.

\subsubsection{Prompt Engineering}
\label{subsubsection:Prompt Engineering}
To guide LLMs toward structured and accurate profiling, we incorporate Chain-of-Thought (CoT) reasoning(\cite{cot}). By compelling the model to articulate its explicit reasoning steps, this approach enhances interpretability while simultaneously reducing hallucinations. We impose a conciseness requirement on the CoT process to prevent overthinking and redundancy. Additionally, we designed task-specific constraints aimed at resolving particular failure modes. For instance, we clarify that the purchase of ``renter-friendly'' items does not categorize a user as a renter, as homeowners may also prefer non-destructive solutions. Beyond improving accuracy, the task-specific prompt effectively minimizes the computational overhead required. A unified abstraction of our prompt template is presented below, and the full implementation is detailed in Appendix \ref{appendix:prompt} for further reference.

\definecolor{prompt_bg}{RGB}{252, 248, 245}  
\definecolor{prompt_frame}{RGB}{191, 97, 106}  
\definecolor{text_emphasis}{RGB}{180, 50, 60}   
\definecolor{text_info}{RGB}{0, 100, 150}        
\begin{tcolorbox}[
    enhanced,
    colback=prompt_bg, 
    colframe=prompt_frame,      
    fonttitle=\bfseries\large\sffamily,
    title={User Profile Prompt Template},
    attach boxed title to top center={yshift=-2mm},
    boxed title style={
        colback=prompt_frame, 
        colframe=prompt_frame,
        arc=4pt,
        boxsep=4pt,
        left=8pt,
        right=8pt,
        coltitle=white, 
    },
    arc=4pt,
    boxrule=1pt,
    left=12pt,
    right=12pt,
    top=10pt,
    bottom=10pt,
    before upper={\setlength{\parindent}{0pt}},  
    coltext=black,                               
]
\vspace{2pt}
\SectionTag{RoleColor}{\# Role} \\
You are a professional user profile analyst. Given a user's behavioral data, provide accurate profile attributes with concise reasoning steps.
\vspace{8pt}
 \\
\SectionTag{MandatoryColor}{\# Task} \\
\textbf{Objective:} \textcolor{red!70!black}{\{Define task-specific profiling objectives\}} \\
\textbf{Constraint:} \textcolor{red!70!black}{\{Specify task-specific constraints to mitigate particular failure modes\}}
\vspace{8pt}
 \\
\SectionTag{InputColor!50}{\# Input} \\
\textbf{Behavioral Data:} \textcolor{blue!70!black}{\{Insert multi-source user behavior data\}} 
\vspace{8pt}
 \\
\SectionTag{OutputColor}{\# Output} \\
\textbf{Format:} \textcolor{yellow!90!black}{\{Return valid JSON with concise reasoning and profile attributes\}} 
\vspace{8pt}
\end{tcolorbox}

\subsubsection{Data Preparation}
\label{subsubsec:Data Preparation}
\subsubsection*{Legacy Heuristic Label}
Although legacy heuristic labels may suffer from limited coverage and logical conflicts, they serve as pseudo ground truth during our data synthesis process.  We define a rule-based labeling function, denoted by $\mathrm{RULE}(\cdot)$, to infer atomic pseudo-labels $\tilde{\mathbf{y}}^a$ from the user behavioral corpus $U$. For example, if a user mentions a spouse or has children in a comment, they are labeled as married. Conversely, if comments mention girl/boy friends or if the user is a minor, they are considered unmarried. This rule-based annotation process is formally expressed in Eq. (\ref{eq:legacy heuristic label}):

\begin{equation}
\tilde{\mathbf{y}}^a= \mathrm{RULE}(U),
\label{eq:legacy heuristic label}
\end{equation}

After annotating all users, we apply stratified sampling to ensure a balanced demographic distribution, yielding an initial candidate set $Q_{\text{init}}$. Each question $q \in Q_{\text{init}}$ is an attribute-specific synthesis prompt constructed from a user’s behavioral corpus, designed as the input to the teacher LLM for generating supervision data for a particular atomic attribute.
\vspace{-0.3cm}
\subsubsection*{Low-Entropy Data Filtering}
Before Supervised Fine-Tuning (SFT), we perform a data curation step to filter low-entropy questions and improve training efficiency. We use a LLM (specifically Qwen3-8B Thinking) to evaluate the initial candidate set $Q_{\text{init}}$ of 1 million questions. We discard questions where the model's prediction perfectly matches legacy heuristic labels, as these ``easy'' questions offer minimal gradient utility. This curation step is formalized as Eq. (\ref{eq:low-entropy data filtering}), refining $Q_{\text{init}}$ to 490,000 questions:
\begin{equation}
Q_{\text{filtered}} = \{q \in  Q_{\text{init}} \mid \mathrm{LLM}(q) \neq \tilde{y}^a_q\},
\label{eq:low-entropy data filtering}
\end{equation}
where $Q_{\text{filtered}}$ denotes the filtered dataset, and $\tilde{y}^a_q \in \tilde{\mathbf{y}}^a$ represents the heuristic label associated with question $q$.

\vspace{-0.3cm}
\subsubsection{Multi-Stage SFT}
\label{subsec:Multi-Stage SFT}
We implement a multi-stage SFT consisting of three stages that progressively increases task complexity, aligning data difficulty with the evolving capability of the model.

\subsubsection*{SFT Data}
\paragraph{Stage 1: Establishing Output Norms via Standard Questions}

\begin{description}[leftmargin=0em,labelsep=1em,itemsep=0.5em,parsep=0pt,topsep=-1pt,
    partopsep=0pt]
    \item[- \textit{Objective}] To calibrate the model’s structural output capabilities and establish foundational proficiency in extracting ``atomic user attributes'' from explicit historical behaviors.

    \item[- \textit{Method}] We utilize two heterogeneous 30B-scale models (Qwen3-30B-A3B-Thinking-2507 and DeepSeek-R1-Distill-Qwen-32B) to generate five independent synthesis outputs $\mathbf{o}_m = \{o_1, \dots, o_5\}$ for each question in $Q_{\text{filtered}}$. To ensure high data fidelity, we then implement a rigorous dual-verification mechanism, which is formally defined in Eq. (\ref{eq:data for stage1-1}-\ref{eq:data for stage1-2}). Specifically, as shown in Eq. (\ref{eq:data for stage1-1}), we first use an atomic extraction operator $\mathcal{F}_a$ to derive the attribute label $\hat{y}^a_{o_i}$ from each output $o_i$. Then, as detailed in Eq. (\ref{eq:data for stage1-2}), a sample $(q, o_\text{rand})$ is accepted and included in our final dataset, $D_{\text{stage1}}$, only if at least four of the five extracted labels are consistent with the original pseudo-label $\tilde{y}^a_q$.
    \begin{align}
    \hat{y}^a_{o_i} &= \mathcal{F}_a(o_i), \quad \forall i \in \{1, \dots, 5\} \label{eq:data for stage1-1}\\
    D_{\text{stage1}} &= \left\{ (q, o_\text{rand}) \mid q \in Q_{\text{filtered}},\sum_{i=1}^5 \mathbb{1}(\hat{y}^a_{o_i} = \tilde{y}^a_q) \geq 4  \right\},
    \label{eq:data for stage1-2}
    \end{align}

    where $\mathbb{1}(\cdot)$ is the indicator function and each final data point pairs the question $q$ with $o_\text{rand}$, a randomly selected correct output from the verified set.
    
    \item[- \textit{Outcome}] This entire process yielded an initial set of approximately 210,000 verified questions, which was then sampled to form the final collection of approximately 130,000 high-quality samples in $D_{\text{stage1}}$, ensuring a balanced feature distribution for efficient training.
\end{description}

\vspace{-0.3cm}
\paragraph{Stage 2: Enhancing Robustness via Controversial Questions}
\begin{description}[leftmargin=0em,labelsep=1em,itemsep=0.5em,parsep=0pt,topsep=-1pt,
    partopsep=0pt]
    \item[- \textit{Objective}] To enhance model robustness in handling boundary cases, long-tail distributions, and ambiguous behavioral patterns, thereby enabling the inference of ``atomic user attributes'' from implicit historical behaviors.
    \item[- \textit{Method}] Our method for this stage involves a two-step process: first, identifying ``Controversial Questions'' that represent high-variance regions of the data manifold, and second, resynthesizing a high-fidelity output for them using a more powerful model (Qwen3-235B-A22B-Thinking-2507) , guided by optimized prompts to accurately model the “atomic user attributes”. The entire selection and verification process for constructing the final dataset, $D_{\text{stage2}}$, is formally captured in Eq. (\ref{eq:data for stage2}). Eq. (\ref{eq:data for stage2}) enforces two critical conditions. First, the criterion $0 < \sum_{i=1}^5 \mathbb{1}(\hat{y}^a_{o_i} = \tilde{y}^a_q) < 4 $ mathematically defines a "Controversial Question": it selects questions that showed inter-model disagreement (i.e., fewer than four agreed), but where at least one output is in alignment with legacy heuristic labels, ensuring a valid signal exists. Second, the condition $\mathcal{F}_a(o_r) = \tilde{y}^a_q$ serves as a verification step, ensuring that the new, resynthesized output $o_r$ is consistent with the legacy heuristic label.
    \begin{equation}
    D_{\text{stage2}} = \left\{ (q, o_r) \mid q \in Q_{\text{filtered}}, 0 < \sum_{i=1}^5 \mathbb{1}(\hat{y}^a_{o_i} = \tilde{y}^a_q) < 4 , \mathcal{F}_a(o_r) = \tilde{y}^a_q \right\}.
    \label{eq:data for stage2}
    \end{equation}

    \item[- \textit{Outcome}] This process yielded a set of approximately 250,000 verified ``Controversial Questions'', which was subsequently filtered for feature balance and training efficiency, resulting in a final collection of approximately 170,000 samples, denoted as $D_{\text{stage2}}$.
\end{description}
\vspace{-0.3cm}
\paragraph{Stage 3: Composite Profile Summary via Hard Questions}
\begin{description}[leftmargin=0em,labelsep=1em,itemsep=0.5em,parsep=0pt,topsep=-1pt,
    partopsep=0pt]
    \item[- \textit{Objective}] To transition from ``atomic user attributes'' to a ``composite user profile summary'', facilitating long-term lifecycle understanding, minimizing computational overhead during production inference, and providing a foundation for AI-native components like the ``Pluggable Memory Module''.
    \item[- \textit{Method}] To capture long-term user evolution, the model's context window was extended from 16K to 36K tokens, enabling the processing of up to a decade of behavioral history. We selected approximately 130,000 seed users with logically consistent heuristic labels across diverse categories. To reduce the complexity of synthesizing composite profiling, we utilized the Qwen3-235B-A22B-Thinking-2507 model, incorporating atomic attributes from the Stage 2 model as intermediate reference signals. This procedure generates a ``composite user profile summary'' $o_c$. Finally, to guarantee high fidelity, a verification step was implemented to retain only those samples where all atomic attributes within $o_c$ maintained strict alignment with the legacy heuristic labels. The construction of the final training dataset, $D_{\text{stage3}}$, is formally captured in Eq. (\ref{eq:data for stage3}).
    \begin{equation}
        D_{\text{stage3}} = \left\{ (q, o_c) \mid q \in Q_{\text{hard}}, \mathcal{F}_c(o_c) = \tilde{\mathbf{y}}^a \right\} \cup D'_{\text{stage2}},
    \label{eq:data for stage3}
    \end{equation}
    
    where $\mathcal{F}_c$ represents the composite extraction operator that retrieves all atomic attributes from the profiling $o_c$, and $\tilde{\mathcal{y}}^a$ denotes all legacy heuristic labels assigned to the user. The term $D'_{\text{stage2}}$ denotes a representative subset from Stage 2, included to prevent catastrophic forgetting of atomic-level tasks.
    \item[- \textit{Outcome}] This stage produced approximately 37,000 high-quality ``Hard Questions''. To preserve atomic profiling proficiency and task diversity, these were integrated with samples from Stage 2, resulting in a final training dataset of approximately 73,000 samples, denoted as $D_{\text{stage3}}$.
\end{description}

\subsubsection*{SFT Hyperparameters}
We detail the training hyperparameters and configurations across all three SFT stages in Table~\ref{tab:sft_hyperparams}.

\begin{table}[htbp]
  \centering
  \caption{Hyperparameter configurations for the three SFT stages.}
  \label{tab:sft_hyperparams}
  \renewcommand{\arraystretch}{1.2} 
  \begin{tabularx}{\textwidth}{l *{3}{>{\centering\arraybackslash}X}}
    \toprule
    \textbf{Hyperparameters} & \textbf{SFT Stage 1} & \textbf{SFT Stage 2} & \textbf{SFT Stage 3} \\
    \midrule
    Learning rate & $2.0 \times 10^{-5}$ & $2.0 \times 10^{-5}$ & $2.0 \times 10^{-5}$ \\
    Batch size & \multicolumn{3}{c}{256 (\#device=32, batch size per device=2, gradient accumulation=4)} \\
    LR scheduler & Cosine & Cosine & Cosine \\
    Gradient norm clip & 1.0   & 1.0   & 1.0 \\
    Optimizer & \multicolumn{3}{c}{AdamW ($\beta_1=0.9$, $\beta_2=0.999$, $\epsilon=1.0 \times 10^{-8}$)} \\
    Use BF16 & TRUE  & TRUE  & TRUE \\
    Max sequence length & 16K   & 16K   & 36K \\
    Training steps & 485   & 658   & 391 \\
    \bottomrule
  \end{tabularx}
\end{table}

\subsubsection{Reinforcement Learning (RL) }
\label{subsubsec:rl}
Reinforcement learning (RL) has proven effective in enhancing the reasoning abilities of language models across diverse reasoning tasks(\cite{Guo_2025,qwen3technicalreport,wang2025reinforcement,zha2025rl}). In this stage, we employ RL to further align the model with composite user profile summary.

\subsubsection*{RL Data}
The RL training dataset is derived from the composite profile summary synthesized in Stage 3. To ensure data quality, these profiles first undergo a rigorous filtering process, which includes verification against legacy heuristic labels and strict deduplication against the SFT dataset. From this refined pool, we then employ a stratified sampling strategy to ensure data diversity, selecting 100 samples from each gender and age group. This procedure yields the RL training dataset $D_{\text{RL}}$ of 1,829 samples.

\subsubsection*{Learning Objective}
We adopt an enhanced version of Group Relative Policy Optimization (GRPO) (\cite{shao2024deepseekmath}). For each prompt $q$, the algorithm samples a group of responses $\{o_1, o_2, \dots, o_G\}$ from the old policy $\pi_{\theta_{\text{old}}}$. The policy $\pi_\theta$ is updated by maximizing the following objective:

\begin{equation}
\begin{aligned}
\mathcal{J}_{\text{GRPO}}(\theta) &= \mathbb{E}\left[ q \sim D_{\text{RL}}, \{o_i\}_{i=1}^G \sim \pi_{\theta_{\text{old}}}(\cdot|q) \right] \\
& \frac{1}{G} \sum_{i=1}^{G} \frac{1}{|o_i|} \sum_{t=1}^{|o_i|} \left\{ 
\min\left[ 
\frac{\pi_\theta(o_{i}|q)}{\pi_{\theta_{\text{old}}}(o_{i}|q)} A_{i,t}, 
\text{clip}\left( 
\frac{\pi_\theta(o_{i}|q)}{\pi_{\theta_{\text{old}}}(o_{i}|q)}, 
1 - \varepsilon, 1 + \varepsilon 
\right) A_{i,t}
\right] 
- \beta \mathbb{D}_{KL} \left[ \pi_\theta \| \pi_{\text{ref}} \right] 
\right\},
\end{aligned}
\end{equation}

where $\varepsilon$ and $\beta$ are hyperparameters. The advantage $A_i$ is computed by normalizing the rewards $\{r_1, r_2, \dots, r_G\}$ within each group:
\begin{equation}
A_i = \frac{r_i - \text{mean}(\{r_k\}_{k=1}^G)}{\text{std}(\{r_k\}_{k=1}^G)}.
\end{equation}
The term $\mathbb{D}_{KL} \left[ \pi_\theta \| \pi_{\text{ref}} \right] $ represents the Kullback-Leibler divergence between the current policy $\pi_\theta$ and a reference model $\pi_{\text{ref}}$ (typically the initial SFT model). This regularization term, scaled by $\beta$, constrains the policy update to prevent it from deviating too far from the base distribution, thereby ensuring training stability and mitigating reward hacking. This approach eliminates the need for an additional critic model, significantly reducing computational overhead while maintaining stable policy updates.

\subsubsection*{Reward Design}
The reward function comprises three components:
\begin{itemize}
    \item \textbf{Format Reward}: Assess whether the output strictly follows the structured format, such as the reasoning process being enclosed within the specified ``<think>'' and ``</think>'', and the final result being enclosed within the specified ``<answer>'' and ``</answer>''. Also, ensure that the content of the final result is valid JSON. Provide a binary score (1 or 0).
    \item \textbf{Atomic Accuracy Reward}: Measures the precision of atomic attributes compared to the ground truth (GT). For discrete attributes (such as gender), accuracy can be measured directly against the GT. For generated attributes (such as interests and preferences), we use the GTE representation model (\cite{li2023general}) to assess similarity with the GT. The score is calculated as the ratio of correctly matched labels to the total number of atomic attributes, preventing this component from dominating the total reward.
    \item \textbf{Summary Quality Reward}: We utilize a powerful thinking model as a judge to evaluate the generated summary based on four dimensions: completeness, consistency, conciseness, and aesthetics. The final summary reward is a weighted average of these scores.
\end{itemize}

\subsubsection*{Dual-Filter Group Relative Policy Optimization (DF-GRPO)}
To enhance training stability and data efficiency, we propose DF-GRPO, a refined strategy that selectively curates training samples rather than treating all generations equally. DF-GRPO implements a hierarchical two-tier filtering mechanism to ensure that the model learns only from high-quality, informative outputs:
\begin{enumerate}
    \item \textbf{Sample-Level Filtering}: Samples are discarded if they reach the maximum sequence length (indicating potential truncation) or fail the \textbf{Format Reward} (indicating a lack of structural integrity). This prevents the model from learning from malformed outputs which could degrade its core profiling capabilities.
    \item \textbf{Group-Level Filtering}: To address the limitations of relative advantages, we introduce global reward thresholds defined by two hyperparameters, $\varepsilon_{\text{low}}$ and $\varepsilon_{\text{high}}$.Groups are excluded from the current update if their average reward falls below $\varepsilon_{\text{low}}$ (indicating low quality) or exceeds $\varepsilon_{\text{high}}$ (indicating prior optimization), thereby avoiding suboptimal gradients or over-optimization.
\end{enumerate}
For the remaining effective samples, the total reward is the sum of the Atomic Accuracy Reward and Summary Quality Reward, as the format requirements are implicitly satisfied through the filtering process.

\subsubsection*{RL Hyperparameters}
Our RL training was conducted with a total batch size of 128. We performed one gradient update per iteration with a learning rate of $1 \times 10^{-6}$. To accommodate complex reasoning, the maximum sequence length was set to 40,000 tokens. During the sampling phase, we used a temperature of 0.7 and a top-$p$ value of 0.9 to ensure sufficient exploration and diversity in the generated responses.

\section{Holistic Persona Reasoning Bench}
\label{sec:HPR-bench}

In the burgeoning era of AI Agents, a model's ability to maintain and reason over a "memory" of the user is paramount for delivering personalized, stateful interactions. However, evaluating whether this memory can effectively summarize a user's holistic persona and infer their evolving intent remains a critical yet unaddressed blind spot in LLM evaluation. While existing benchmarks focus on general-purpose capabilities like code and math, the deep, personalized reasoning fundamental to a useful agent memory lacks systematic assessment.

To fill this critical void, we introduce the \textbf{Holistic Persona Reasoning Bench (HPR-Bench)}, a comprehensive benchmark designed specifically to measure this vital capability. It comprises two key parts: atomic attribute inference and composite profile summarization. The data in HPR-Bench derives from the simulated data mentioned in Section \ref{sec:data}.

\subsection{Task Definition}
We define the task as generating a multi-dimensional user profile from long-term behavioral data. A comprehensive profile comprises several key dimensions, which we conceptualize as atomic attributes denoted by \textbf{atomic portrait tags}. These include:

\begin{itemize}
\item \textbf{Life Stage:} General life-phase categorization (e.g., single, in a relationship, family-oriented).
\item \textbf{Household Context:} Number, age, and gender of children.
\item \textbf{Lifestyle Indicators:} General consumption tier and lifestyle patterns.
\item \textbf{Educational \& Professional Background:} Student status, occupation.
\item \textbf{Geographic Context:} Resident city tier and regional preference.
\end{itemize}

Beyond the accuracy of above individual attributes, we target a more challenging task, \textbf{user profiling summarization}, demanding the model to generate a fluent and holistic summary of a user. 

\subsection{Data Curation}
\begin{figure}[htbp]
    \centering 
    \includegraphics[width=0.95\textwidth]{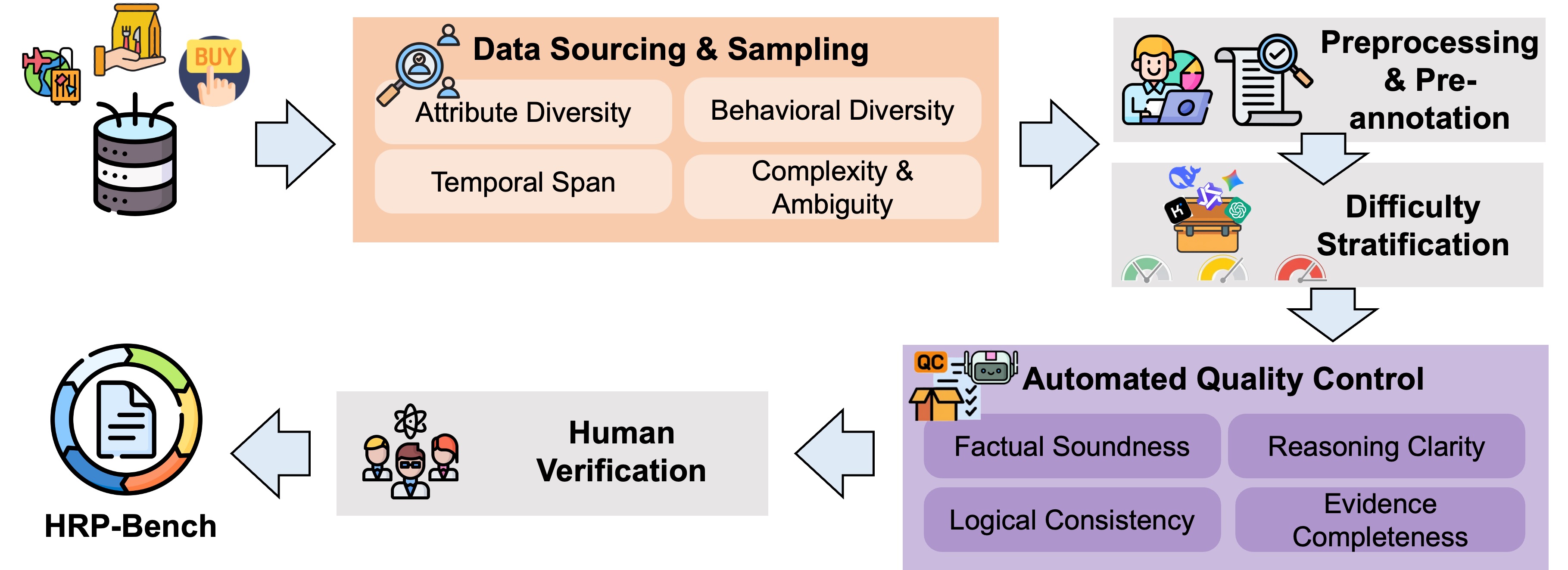}
    \caption{The data curation pipeline of the proposed HPR-Bench, including five steps: (1) Data Sourcing and Sampling. (2) Preprocessing and Pre-annotation. (3) Difficulty Stratification via Model-based Annotation. (4) Automated Quality Control. (5) Human Verification.} 
    \label{fig:HPR-Bench curation} 
\end{figure}

As illustrated in Figure \ref{fig:HPR-Bench curation}, the construction of HPR-Bench follows a multi-stage pipeline to ensure diversity, quality and realistic complexity.

\vspace{-1cm}
\paragraph{(1) Data Sourcing and Sampling} We construct a candidate user pool from a large-scale, multi-sources data hosting. To guarantee data diversity and representativeness, we sample a seed user pool based on the following criteria:

\vspace{-0.5cm}
\begin{itemize}
\item \textbf{Attribute Diversity.} The user pool captures a wide range of ages, genders, and life stages, reflecting diverse behavioral patterns.
\item \textbf{Behavioral Diversity.} Users must be active across multiple sources, including e-commerce (search, clicks, purchases), food delivery (search, purchases), etc.
\item \textbf{Temporal Span.} To capture significant life-stage transitions, we focus on users with sufficiently long behavioral histories to reflect meaningful persona evolution.
\item \textbf{Complexity and Ambiguity.} To evaluate the model's robustness against hallucinations, we deliberately select users whose behaviors are complex and ambiguous, allowing for a stratified difficulty across the test set.
\end{itemize}

\paragraph{(2) Preprocessing and Pre-annotation} First, the raw behavioral data undergoes a manual review to filter out sensitive content and confirm its usability. We then generate a set of reference atomic attributes for each user to serve as a preliminary ground truth. 

\vspace{-0.5cm}
\paragraph{(3) Difficulty Stratification via Model-based Annotation} To stratify the benchmark by difficulty, we first employ a suite of existing LLMs to perform preliminary annotations on the processed data. Difficulty levels are then determined based on inter-model consensus, conflicts among predictions, and consistency with the pre-annotated reference labels.

\vspace{-0.5cm}
\paragraph{(4) Automated Quality Control} A key challenge in user profiling is the model's tendency to over-infer from weak evidence. To mitigate this, we introduce an automated quality control mechanism using frontier models to score the reasoning process of the annotation models with the pre-annotated labels. This evaluation considers factual soundness, clarity of reasoning, logical consistency, and completeness of evidence. This mechanism effectively filters low-quality instances and identifies challenging samples to increase the benchmark's difficulty.

\vspace{-0.5cm}
\paragraph{(5) Human Verification} All instances in the final benchmark undergo a rigorous human verification process. Each sample is independently reviewed by at least five experts skilled at user profiling. A majority voting principle is applied to resolve disagreements. Samples deemed ambiguous or unidentifiable by one or more experts are discarded to ensure the benchmark is human-verifiable. This strict verification standard guarantees a benchmark of exceptional reliability and clarity.

\subsection{Data Distribution}
\begin{figure}[htbp]
    \centering 
    \includegraphics[width=0.95\textwidth]{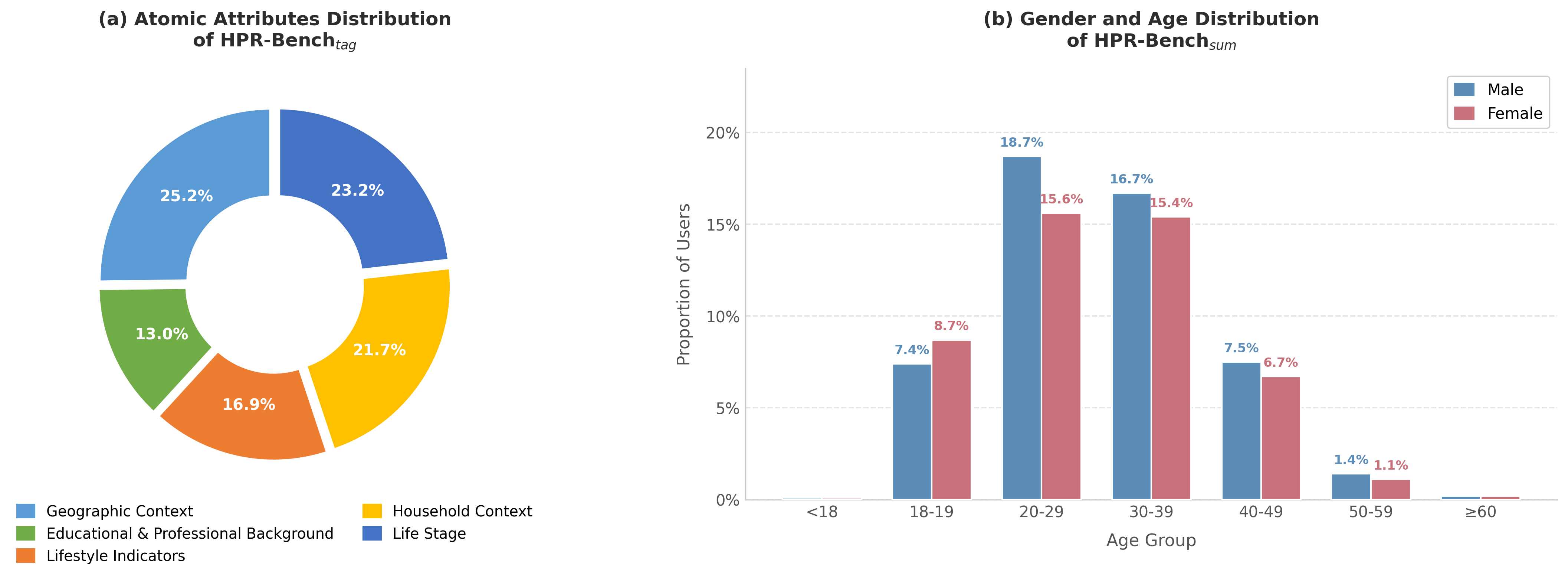}
    \caption{\centering Data Distribution of HPR-Bench}
    \label{fig: data distribution of HPR-Bench}
\end{figure}

The final HPR-Bench benchmark comprises two primary task formats: Atomic Portrait Tag Infer (HPR-Bench\textsubscript{tag}, presented as single-choice or multiple-choice questions) and Profile Summarization (HPR-Bench\textsubscript{sum}, free-form text generation). The tasks cover the full spectrum of defined atomic attributes, providing a robust framework for a fine-grained and holistic evaluation of LLMs in user profiling. The quantity data distribution is described in Figure \ref{fig: data distribution of HPR-Bench}.

\section{Evaluations}
\label{sec:Evaluations}

\subsection{Experimental Setup}
\paragraph{Implementation Details} 
To strike a balance between performance and efficiency, we use Qwen3-8B (\cite{qwen3technicalreport}) as the backbone. For multi-stage curriculum SFT, we use an in-house large language model fine-tuning framework adapted from LLaMA Factory (\cite{zheng2024llamafactory}) with DeepSpeed (\cite{samyam2019zero}) training backend, Flash-attention-2 and bfloat16 precision. For RL training, we use ROLL: Reinforcement Learning Optimization for Large-Scale Learning (\cite{wang2025reinforcement}). The training datasets are described in \cref{subsec:Post-Training}. 

\vspace{-0.5cm}
\paragraph{Benchmarks and Evaluation Metrics}
For in-domain user profiling capabilities evaluation, we use the proposed HPR-Bench benchmark introduced in \cref{sec:HPR-bench}, consisting of atomic portrait tag infer ($\text{HPR-Bench}_\text{tag}$) and profile summarization ($\text{HPR-Bench}_\text{sum}$). To assess the model's generalization capabilities, we also evaluate its performance on several widely-used benchmarks from both general and E-commerce domains.

\textbf{(1) Evaluation of in-domain atomic portrait tag inference}. We evaluate this capability on the proposed $\text{HPR-Bench}_\text{tag}$. To account for generative randomness, we report $\textbf{Avg@k}$ across all 18 atomic tags from five dimensions: Life Stage (LS), Household Context (HC), Lifestyle Indicators (LI), Educational \& Professional Background (EP), and Geographic Context (GC).  $Avg@k$ is calculated by conducting $n$ independent inference trials and counting the average number of correct outcomes. k is set to 10 in this paper.
 
\textbf{(2) Evaluation of in-domain profile summarization}. We perform comprehensive evaluations on $\text{HPR-Bench}_\text{sum}$, encompassing both quantitative metrics and qualitative assessments.
   \begin{itemize}

        \item \textit{Quantitative evaluation}. For the natural language summary, the evaluation is conducted for both grounding and generation quality. Grounding is measured by the accuracy ($\mathbf{Acc_{Ex}}$) and coverage ratio ($\mathbf{COV_{Ex}}$) of atomic tags extracted from the summary via Qwen3-32B(\cite{qwen3technicalreport}). Generation Quality is measured against a reference summary using \textbf{BLEU} (\cite{papineni-etal-2002-bleu}) and semantic similarity ($\mathbf{Score_{sim}}$).

        \item \textit{Qualitative evaluation}. We employ an LLM-as-Judge framework where DeepSeek-V3.2 (\cite{deepseekai2025deepseekv32pushingfrontieropen}) scores each summary from 0 to 10 across four rigorously defined dimensions: (1) \textbf{Completeness \& Granularity}: Assesses whether the summary covers all non-unknown atomic tags while preserving their finest-grained level of detail (e.g., distinguishing between "Teacher" and the more specific "Primary 
        School Teacher"). (2) \textbf{Self-Consistency}: Verifies the summary's self-consistency and plausibility against common sense, ensuring it is free of internal logical contradictions, independent of the ground truth labels. (3) \textbf{Conciseness \& Hallucination}: Measures if the summary expresses all details with maximal information density, avoiding both redundancy and factual hallucination. (4) \textbf{Aesthetic Quality}: Judges the summary's professional tone and natural fluency, evaluating its ability to integrate data into a cohesive narrative.
        
    \end{itemize}

\textbf{(3) Out-of-domain evaluation}. For general capabilities evaluation, we conduct experiments on several widely adopted general domain benchmarks, including C-Eval (\cite{huang2023ceval}), CMNLI, BUSTM (\cite{BUSTM}), $\text{C}^\text{3}$ (\cite{sun2019investigatingpriorknowledgechallenging}). For the e-commerce task, we utilize the subset of ChineseEcomQA benchmark (\cite{chen2025chineseecomqascalableecommerceconcept}), including three core concepts: Relevance Concept (RLC), Review Concept (RVC) and Category Concept (CC). Performance across all benchmarks are measured using $\mathbf{Avg@k}$.

\subsection{Main results}
\label{sec:main_results}

\subsubsection{In-domain Experimental Results}
To evaluate user profiling capabilities, we benchmark our model against three categories of baselines. First, for traditional methods, we assess a standard discriminative prediction model (DNN, \cite{LURM}), which is widely deployed in current production systems. Second, we compare our approach with state-of-the-art (SOTA) Large Language Models (LLMs), categorized into ``Thinking'' and ``Non-Thinking'' modes. This suite includes Qwen3-8B (Thinking/Non-Thinking), Qwen3-235B-A22B-Thinking/Instruct-2507 (\cite{qwen3technicalreport}), Qwen3.6-Plus (\cite{qwen36plus}), Kimi-K2.5 (\cite{kimi2026kimiK2.5}), GLM-5 (\cite{glm5team2026glm5vibecodingagentic}), DeepSeek-V3.2 (\cite{deepseekai2025deepseekv32pushingfrontieropen}), and DeepSeek-R1 (\cite{Guo_2025}). The comprehensive results are presented in Table \ref{tab:overall_tag} for $\text{HPR-Bench}_\text{tag}$ and Table \ref{tab:overall_summary} for $\text{HPR-Bench}_\text{sum}$. More detailed results refer to Appendix \ref{appendix:Additional Evaluation Results}.

\begin{table}[htbp]
  \centering
  \caption{Performance on $\text{HPR-Bench}_\text{tag}$, including overall Avg score and five atomic attributes dimensions Avg@10 score (LS: Life Stage, HC: Household Context, LI: Lifestyle Indicators, EP:Educational \& Professional Background, GC: Geographic Context).}
    \scriptsize
    \begin{tabular}{p{16em}p{10em}cccccc}
    \toprule
    \multirow{2}[4]{*}{\textbf{Models}} & \multirow{2}[4]{*}{\textbf{\# Params}} & \multicolumn{1}{c}{\multirow{2}[4]{*}{\textbf{Avg}}} & \multicolumn{1}{p{2.915em}}{\textbf{LS}} & \multicolumn{1}{p{3.25em}}{\textbf{HC}} & \multicolumn{1}{p{3.335em}}{\textbf{LI}} & \multicolumn{1}{p{3.165em}}{\textbf{EP}} & \multicolumn{1}{p{3.085em}}{\textbf{GC}} \\
\cmidrule{4-8}    \multicolumn{1}{l}{} & \multicolumn{1}{l}{} &       & \multicolumn{1}{p{2.915em}}{\textbf{Avg@10}} & \multicolumn{1}{p{3.25em}}{\textbf{Avg@10}} & \multicolumn{1}{p{3.335em}}{\textbf{Avg@10}} & \multicolumn{1}{p{3.165em}}{\textbf{Avg@10}} & \multicolumn{1}{p{3.085em}}{\textbf{Avg@10}} \\

    \midrule
    \rowcolor{gray!10} \multicolumn{8}{c}{Discriminative Models} \\
    \midrule
    DNN  & \multicolumn{1}{l}{} & 0.5846  & 0.7156  & 0.4000   & 0.6525  & 0.5616  & \multicolumn{1}{p{3.085em}}{—} \\
    \midrule
    \rowcolor{gray!10} \multicolumn{8}{c}{Non-Thinking Models} \\
    \midrule
    Qwen3-8B Non-Thinking & Dense (8B) & 0.4612  & 0.5807  & 0.3281  & 0.6091  & 0.4282  & 0.4130  \\
    Qwen3-235B-A22B-Instruct-2507 & MoE (235B/A22B) & 0.5244  & 0.5944  & 0.4140  & 0.5194  & 0.5584  & 0.5341  \\
    
    \midrule
    \rowcolor{gray!10} \multicolumn{8}{c}{Thinking Models} \\
    \midrule
    Qwen3-8B Thinking & Dense (8B) & 0.5035  & 0.6224  & 0.3831  & 0.6244  & 0.4770  & 0.4599  \\
    Qwen3-235B-A22B-Thinking-2507 & MoE (235B/A22B) & 0.6434  & 0.6411  & 0.4135  & 0.6438  & 0.6261  & 0.7478  \\
    Kimi-K2.5  & MoE (1T/A32B) & 0.7064 &	0.7467 & 	0.538 &	0.6866 	& 0.659 &	0.7834  \\
    DeepSeek-R1-0528 & MoE (671B/37B) & 0.6613  & 0.6412  & 0.4752  & 0.6664  & 0.6786  & 0.7424  \\
    DeepSeek-V3.2 & MoE (671B/37B) & 0.6684 & 0.6842 & 0.4981 &	0.6753 & 0.6693 & 0.7314 \\
    GLM-5 & MoE (744B/A40B) & 0.7186 & 0.7691 & 0.5345 &	\textcolor[rgb]{ 1,  0,  0}{\textbf{0.7098}}  & 0.7037  & \uline{0.7838} \\

    Qwen3.6-Plus & —  &		\textcolor[rgb]{ 1,  0,  0}{\textbf{0.7329}} &	\textcolor[rgb]{ 1,  0,  0}{\textbf{0.7907}} &	\uline{0.5965} &	0.6991 &	0.6694 &	\textcolor[rgb]{ 1,  0,  0}{\textbf{0.7993}} \\
    
    \midrule
    $\text{UserGPT}_\text{SFT}$ & Dense (8B) & \uline{0.7325}  & 0.7776  & \textcolor[rgb]{ 1,  0,  0}{\textbf{0.6088}}  & \uline{0.6999}  & \textcolor[rgb]{ 1,  0,  0}{\textbf{0.8212}}  & 0.7548  \\
    UserGPT & Dense (8B) & 0.7306 &	\uline{0.7879} &	0.5962  &	0.6945 & 	\uline{0.8164}  &	0.7546  \\
    \bottomrule
    \end{tabular}%
  \label{tab:overall_tag}%
\end{table}%

\vspace{-0.5cm}
\paragraph{Curriculum-Driven Post-Training significantly boosts the persona reasoning of small-sized LLMs.}
On the atomic portrait tag prediction, UserGPT achieves a relative improvement of \textbf{45.10\%}（from 0.5035 to 0.7306) in average Avg@10 over the backbone Qwen3-8B Thinking, with consistent improvements across all five dimensions (Life Stage, Household Context, Lifestyle Indicators, Educational \& Professional Background and Geographic Context). For summary generation, it yields significant relative improvements both quantitatively and qualitatively over the backbone Qwen3-8B Thinking: \textbf{+50.47\%} on $\textbf{Acc}_\textbf{Ex}$, \textbf{+10.75\%} on $\textbf{COV}_\textbf{Ex}$, \textbf{+18.77\%} on total score judge by LLM.
Remarkably, despite its significantly smaller scale, UserGPT achieves performance comparable to or surpassing much larger state-of-the-art models on both tasks. On tag prediction, UserGPT attains an Avg@10 of \uline{0.7325}, nearly matching  Qwen3.6-Plus (\textbf{0.7329}). On summary generation, UserGPT outperforms Qwen3-235B-A22B-Thinking-2507 on key quantitative metrics, with $\text{Acc}_\text{Ex}$ of 
\textbf{0.7528} vs.\ \uline{0.7014} (\textbf{+7.31\%}) and 
$\text{COV}_\text{Ex}$ of \textbf{0.9747} vs.\ \uline{0.9638} (\textbf{+1.13\%}), while remaining competitive on qualitative scores across four dimensions.

\vspace{-0.3cm}
\paragraph{Even the SOTA reasoning models exhibit persistent deficiencies in user persona prediction.}
Despite remarkable progress in general domains, current LLMs exhibit significant deficiencies in user profile prediction, as evidenced in Table \ref{tab:overall_tag}. The best-performing model, Qwen3.6-Plus, achieves only an average accuracy of 0.7329 on the tag prediction task, which further deteriorates to 0.5965 on complex dimensions like Household Context and 0.5713 on $Acc_{\text{Ex}}$ on Profile Summary Generation task. This decline highlights a critical bottleneck: existing models fail to capture the subtle, implicit cues essential for inferring personal attributes. Moreover, we observe negligible performance differences between the ``Thinking'' and ``Non-Thinking'' variants of these advanced LLMs. This suggests that standard chain-of-thought reasoning is insufficient for bridging the gap in user understanding, revealing the inability of current models to handle personalized reasoning without domain-specific adaptation.

In contrast, our approach addresses these limitations through meticulously synthesized data and a curriculum-driven post-training strategy. By constructing high-quality training samples that explicitly target implicit inference patterns (e.g. Controversial Questions in Section \ref{subsec:Multi-Stage SFT}) and organizing them into a progressive learning curriculum, our model effectively internalizes the logic of user profiling. Appendix \ref{appendix:case study} shows a case study. This targeted methodology allows UserGPT to achieve a substantial leap in performance, significantly outperforming existing SOTA models in user understanding.

\begin{table}[!t]
  \centering
  \caption{Quantitative and Qualitative Evaluation results on $\text{HPR-Bench}_\text{sum}$. 
  Quantitative metrics: $\text{Acc}_\text{Ex}$, $\text{COV}_\text{Ex}$, BLEU\_2, BLEU\_4, $\text{Score}_\text{sim}$.
  Qualitative dimensions: Completeness(Comp.), Consistency(Consis.), Conciseness(Concis.), Aesthetics(Aesth.), scoring 0--10 each.}
  \footnotesize
  \setlength{\tabcolsep}{4pt}
  \scalebox{0.90}{
    \begin{tabular}{p{13em}ccccc ccccc}
    \toprule
    \multirow{2}[4]{*}{\textbf{Models}} 
      & \multicolumn{5}{c}{\textbf{Quantitative Evaluation}} 
      & \multicolumn{5}{c}{\textbf{Qualitative Evaluation}} \\
    \cmidrule(lr){2-6} \cmidrule(lr){7-11}
      & \multicolumn{1}{p{3em}}{$\textbf{Acc}_\textbf{Ex}$} 
      & \multicolumn{1}{p{3em}}{$\textbf{COV}_\textbf{Ex}$} 
      & \multicolumn{1}{p{3em}}{\textbf{BLEU\_2}} 
      & \multicolumn{1}{p{3em}}{\textbf{BLEU\_4}} 
      & \multicolumn{1}{p{4em}}{$\textbf{Score}_\textbf{sim}$}
      & \multicolumn{1}{p{5em}}{\textbf{Comp.}} 
      & \multicolumn{1}{p{4.5em}}{\textbf{Consis.}} 
      & \multicolumn{1}{p{4.5em}}{\textbf{Concis.}} 
      & \multicolumn{1}{p{4em}}{\textbf{Aesth.}} \\

    \midrule
    \rowcolor{gray!10} \multicolumn{11}{c}{\textit{Non-Thinking Models}} \\
    \midrule
    Qwen3-8B Non-Thinking 
      & 0.5160 & 0.8821 & 0.0828 & 0.0175 & 0.7811 
      & 4.7568 & 9.7795 & 4.9616 & 4.4236  \\
    Qwen3-235B-A22B-Instruct-2507 
      & 0.6423 & 0.9471 & 0.1616 & 0.0427 & 0.7670 
      & 5.9192 & 9.8210 & 6.0860 & 6.9236  \\

    \midrule
    \rowcolor{gray!10} \multicolumn{11}{c}{\textit{Thinking Models}} \\
    \midrule
    Qwen3-8B Thinking 
      & 0.5192 & 0.8801 & 0.0817 & 0.0153 & 0.7732 
      & 4.7062 & 9.8515 & 4.8489 & 4.9192  \\
    Qwen3-235B-A22B-Thinking-2507 
      & \uline{0.7014} & \uline{0.9638} & \uline{0.1880} & \uline{0.0592} & 0.7917 
      & \uline{6.2549} & 9.8097 & 6.1296 & \uline{7.1770}  \\
    Kimi-K2.5 
      & 0.6356 & 0.9332 & 0.1491 & 0.0318 & 0.7867 
      & 5.6517 & 9.8624 & 5.7825 & 6.4214  \\
    DeepSeek-R1-0528 
      & 0.6255 & 0.9221 & 0.1185 & 0.0232 & 0.7666 
      & 5.6062 & 9.8428 & 5.4998 & 5.3581  \\
    DeepSeek-V3.2 
      & 0.6304 & 0.9341 & 0.1665 & 0.0448 & 0.8062 
      & 5.6963 & 9.8493 & 6.0052 & 5.9607  \\
    GLM-5 
      & 0.6775 & 0.9422 & 0.1804 & 0.0514 & \uline{0.8113} 
      & 6.0432 & 9.8734 & \uline{6.1900} & 6.5677  \\
    Qwen3.6-Plus 
      & 0.5713 & 0.9046 & 0.1371 & 0.0262 & 0.7728 
      & 5.4473 & \uline{9.8839} & 5.5883 
      & \textcolor[rgb]{1,0,0}{\textbf{7.8214}} \\

    \midrule
    \rowcolor{gray!10} \multicolumn{11}{c}{\textit{Our Methods}} \\
    \midrule
    $\text{UserGPT}_\text{SFT}$ 
      & 0.6899 & 0.9587 & 0.1631 & 0.0378 & 0.7854 
      & 6.0447 & 9.8406 & 6.1253 & 6.4236 \\
    UserGPT 
      & \textcolor[rgb]{1,0,0}{\textbf{0.7528}} 
      & \textcolor[rgb]{1,0,0}{\textbf{0.9747}} 
      & \textcolor[rgb]{1,0,0}{\textbf{0.1931}} 
      & \textcolor[rgb]{1,0,0}{\textbf{0.0691}} 
      & \textcolor[rgb]{1,0,0}{\textbf{0.8215}} 
      & \textcolor[rgb]{1,0,0}{\textbf{6.3559}} 
      & \textcolor[rgb]{1,0,0}{\textbf{9.8996}} 
      &  \textcolor[rgb]{1,0,0}{\textbf{6.5882}}  
      & 6.0502 \\
    \bottomrule
    \end{tabular}%
    }
  \label{tab:overall_summary}%
\end{table}

\vspace{-0.5cm}
\paragraph{Moreover, RL further significantly enhances the model's overall tag accuracy, coverage and consistency.}
We conduct experiments on variants of our model, including $\text{UserGPT}_\text{SFT}$ with only multi-stage curriculum SFT, and full UserGPT with both multi-stage curriculum SFT and RL. UserGPT achieves comparable results with $\text{UserGPT}_\text{SFT}$ on the atomic portrait tag prediction task, and presents significantly relative improvements on the summary generation task: \textbf{+9.12\%} on $\text{Acc}_\text{Ex}$, \textbf{+1.67\%} on $\text{COV}_\text{Ex}$, demonstrating the effectiveness of our well-designed RL post-training.

\begin{figure}[htbp]
    \centering 
    \includegraphics[width=0.95\textwidth]{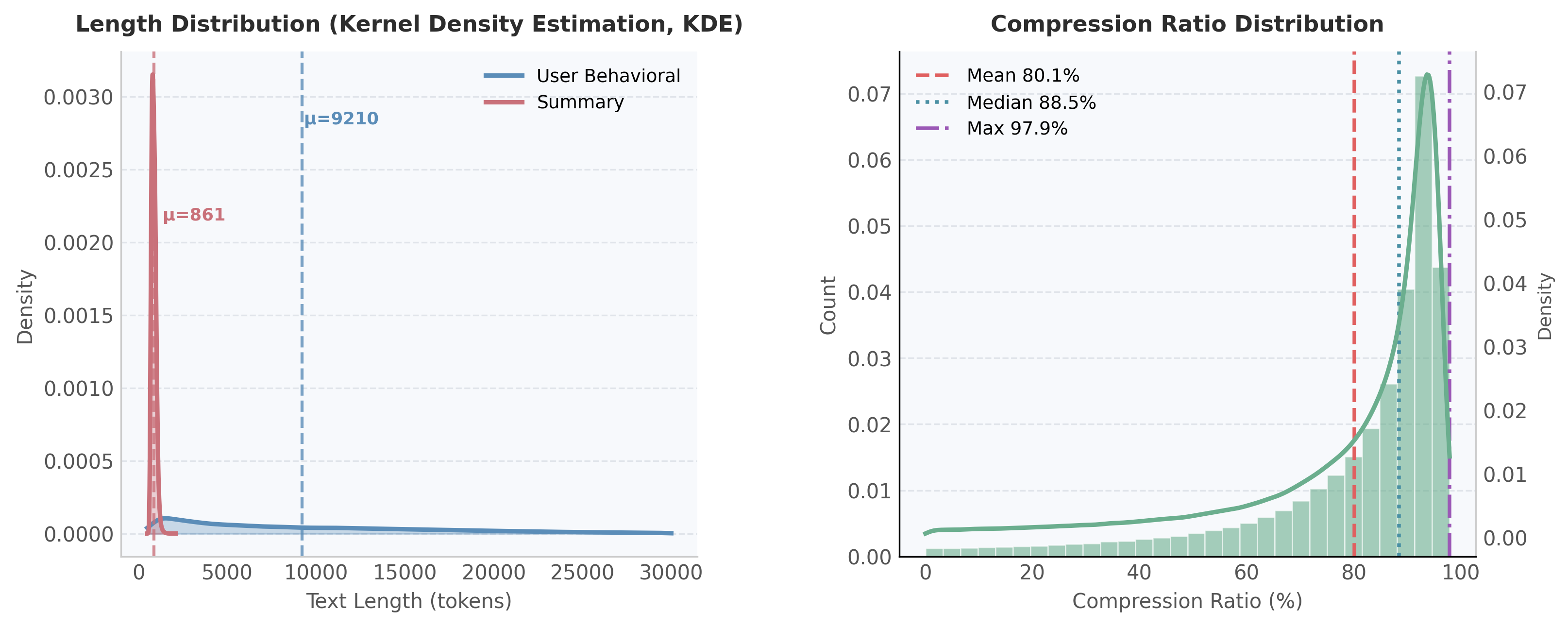}
    \caption{\centering Compression Analysis: User Behavior Context vs. Summary}
    \label{fig:compression}
\end{figure}
\vspace{-1cm}
\paragraph{The generated user profile summaries effectively condense users' long-term behavioral history.} Illustrated in Figure \ref{fig:compression}, we analyzed the length distribution of the original user behavior context and profile summaries generated by UserGPT. The summarization process demonstrates remarkable efficiency, achieving a reduction up to \textbf{97.9\%} in token length, compressing an average of \textbf{15K} input tokens from user history to a concise \textbf{1.2K} tokens summary.

Crucially, this compression does not come at the cost of information integrity. As shown in Table \ref{tab:overall_summary}, the high coverage of core tags extracted from these summaries ($\mathbf{COV_{Ex}\approx 0.9747}$) validates its completeness. This result confirms that our method effectively condenses extensive user histories while preserving critical personal details, showcasing a strong balance between efficiency and fidelity.

\subsubsection{Out-of-Domain Experimental Results}
To evaluate the model robustness after our post-training, we evaluate UserGPT against its backbone model, Qwen3-8B, on out-of-domain benchmarks, including C-Eval, CMNLI, BUSTM, $\text{C}^\text{3}$ and the subset of ChineseEcomQA. As shown in Table \ref{tab:overall_general}, it experiences only a marginal performance drop of 1.32 percentage points compared to the specialized Qwen3-8B Thinking. Crucially, it significantly outperforms the Qwen3-8B Non-Thinking variant by \textbf{5.95} percentage points. This result confirms that our curriculum-driven post-training strategy successfully enhances domain-specific abilities without causing a catastrophic collapse in general-purpose performance.

\begin{table}[htbp]
  \centering
  \caption{On out-of-domain evaluation results between the backbone and UserGPT. We compute the Avg@k on each general domain and E-commerce domain benchmark, as well as the average Avg@k (denoted as Avg) on all those benchmarks.}
  \footnotesize
    \begin{tabular}{p{12em}p{1.8em}p{0.1em}ccccp{0.1em}ccc}
    \toprule 
    \multirow{2}[4]{*}{\textbf{Models}} & \multicolumn{1}{c}{\multirow{2}[4]{*}{\textbf{Avg}}} &       & \multicolumn{4}{c}{\textbf{General domain}} &       & \multicolumn{3}{c}{\textbf{E-commerce domain}} \\
\cmidrule{4-7}\cmidrule{9-11}    \multicolumn{1}{c}{} &          &       & \multicolumn{1}{p{3em}}{\textbf{C-Eval}} & \multicolumn{1}{p{2em}}{\textbf{CMNLI}} & \multicolumn{1}{p{1.8em}}{\textbf{BUSTM}} & \multicolumn{1}{p{2em}}{$\textbf{C}^\textbf{3}$} &       & \multicolumn{1}{p{1.5em}}{\textbf{RLC}} & \multicolumn{1}{p{1.5em}}{\textbf{RVC}} & \multicolumn{1}{p{1.5em}}{\textbf{CC}} \\
\cmidrule{1-3}\cmidrule{4-7}\cmidrule{9-11}    
    Qwen3-8B Thinking  & 0.7302  &       & 0.8271  & 0.5962  & 0.7901  & 0.9384  &       & 0.6036  & 0.6859  & 0.6701  \\
    Qwen3-8B Non-Thinking   & 0.6575  &       & 0.7398  & 0.4492  & 0.7621  & 0.9153  &       & 0.5734  & 0.5634  & 0.5995  \\
    \midrule
    $\text{UserGPT}_\text{SFT}$      & 0.7109  &   &	0.7628 & 0.5726 & 0.7488 & 0.9085  &   & 0.5407 & 0.8000 & 0.6433 \\
    UserGPT  & 0.7170  & & 0.7621 & 0.5755 & 0.7558 & 0.9158 & & 0.5230 & 0.8225 & 0.6644               \\
\bottomrule 
\end{tabular}%
  \label{tab:overall_general}%
\end{table}%
\vspace{-0.5cm}
\subsection{Ablation Study}
In this section, we conduct an in-depth analysis of our training strategies. To ensure a stable and robust evaluation, we employ the \textbf{Pass@k} metric in our ablation studies. For each sample, we perform $n$ independent inference trials and determine the number of correct predictions, denoted by $c$. The Pass@k score is defined as the probability that at least one correct outcome is obtained within a random subset of $k$ trials. In our experimental setup, we set $n = 10$.

\vspace{-0.3cm}
\subsubsection{Data Quality Importance}
\vspace{-0.3cm}
\begin{wrapfigure}{l}{0.48\textwidth}
    \centering
    \includegraphics[width=0.45\textwidth]{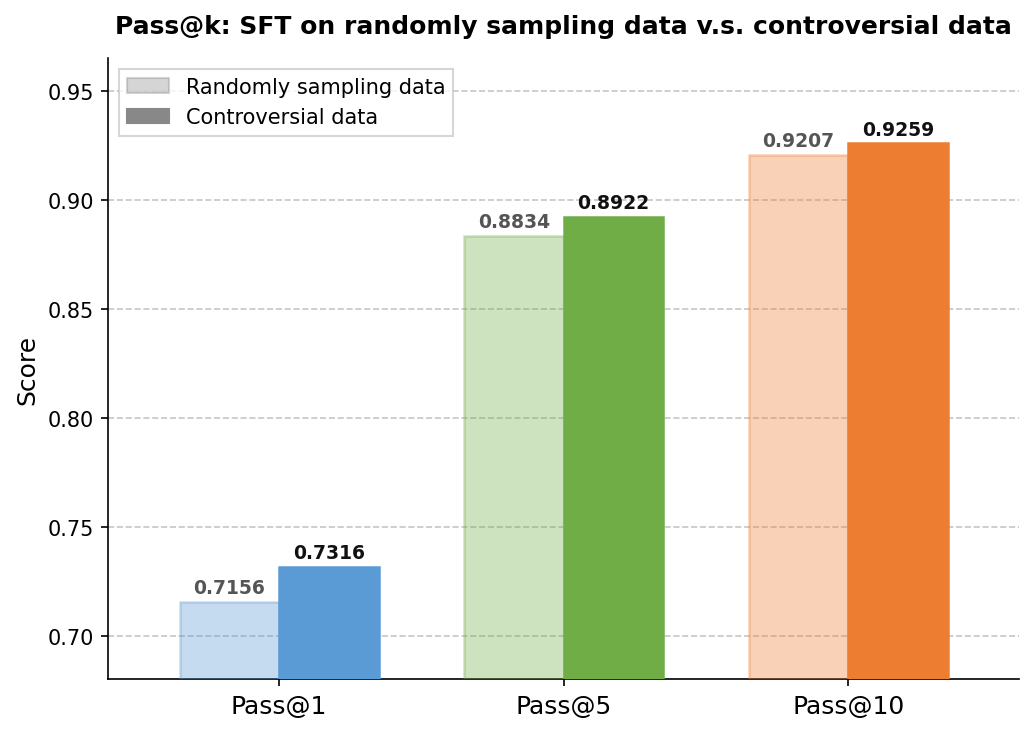}
    \caption{Ablation study regarding the impact of data quality on model performance.}
    \label{fig: ablation on data quality}
\end{wrapfigure}
To investigate the impact of data quality, we conduct an ablation study using Qwen3-8B as the backbone. We compare two models trained on datasets of identical volume: the first is trained on a randomly sampled subset of our full synthetic data, which includes a mixture of easy, standard, and controversial questions; the second is trained exclusively on high-value, challenging samples consisting of controversial questions. 

Performance is evaluated on $\text{HPR-Bench}_\text{tag}$. As illustrated in Figure \ref{fig: ablation on data quality}, training exclusively on high-quality data yields substantially superior results. The most prominent improvement occurs in the Pass@1 score, which reaches 0.7316, indicating a significant increase in the model's confidence when making single-shot predictions on ambiguous samples. Furthermore, the model's theoretical capability, as estimated by Pass@5 and Pass@10 scores, also shows marked gains, reaching 0.8922 and 0.9259 respectively. This stark performance gap underscores that by training on challenging samples, the model is compelled to move beyond simple pattern matching and instead learns to capture the subtle, implicit cues required for reasoning about boundary cases.

\subsubsection{The Necessity of Curriculum Learning.}
\begin{table}[htbp]
  \centering
  \footnotesize
  \caption{The Effectiveness of Multi-stage Curriculum SFT evaluated on $\text{HPR-Bench}_\text{tag}$ and $\text{HPR-Bench}_\text{sum}$.}
    \begin{tabular}{cccccccccccc}
    \toprule
    \multicolumn{4}{c}{\textbf{Training Strategy}} &       & \multicolumn{3}{c}{$\textbf{HPR-Bench}_\textbf{tag}$} &       & \multicolumn{3}{c}{$\textbf{HPR-Bench}_\textbf{sum}$} \\
\cmidrule{1-4}\cmidrule{6-8}\cmidrule{10-12}
\multicolumn{1}{p{2em}}{\textbf{stage1}} & \multicolumn{1}{p{2em}}{\textbf{stage2}} & \multicolumn{1}{p{3.2em}}{\textbf{summary}} & \multicolumn{1}{p{8.7em}}{\textbf{Curriculum-Driven}} &       & \multicolumn{1}{p{1.7em}}{\textbf{Pass@1}} & \multicolumn{1}{p{1.7em}}{\textbf{Pass@5}} & \multicolumn{1}{p{1.7em}}{\textbf{Pass@10}} &       & \multicolumn{1}{p{1.7em}}{$\textbf{Acc}_\textbf{Ex}$} & \multicolumn{1}{p{1.7em}}{$\textbf{COV}_\textbf{Ex}$} & \multicolumn{1}{p{2.2em}}{$\textbf{Score}_\textbf{sim}$} \\
\cmidrule{1-4}\cmidrule{6-8}\cmidrule{10-12}
     &       &       &       &       & 0.5030  & 0.6881  & 0.7385  &       & 0.5192  & 0.8801  & 0.7732  \\ 
     
    \checkmark &       &       &       &       & 0.6996  & 0.8233  & 0.8574  &       & 0.4780  & 0.8247  & 0.7537  \\
    \checkmark & \checkmark &       &       &       & 0.7326  & 0.8881  & 0.9229  &       & 0.4926  & 0.8206  & 0.7567  \\
    \checkmark & \checkmark & \checkmark &       &       & \textcolor[rgb]{ 1,  0,  0}{\textbf{0.7420 }} & 0.8843  & 0.9139  &       & 0.6595  & 0.9563  & 0.7848  \\
    \checkmark & \checkmark & \checkmark & \checkmark &       & 0.7325  & \textcolor[rgb]{ 1,  0,  0}{\textbf{0.8911 }} & \textcolor[rgb]{ 1,  0,  0}{\textbf{0.9230 }} &       & \textcolor[rgb]{ 1,  0,  0}{\textbf{0.6899 }} & \textcolor[rgb]{ 1,  0,  0}{\textbf{0.9587 }} & \textcolor[rgb]{ 1,  0,  0}{\textbf{0.7854 }} \\
    \bottomrule
    \end{tabular}%
  \label{tab:curriculum sft}%
\end{table}%
To deconstruct the contribution of tag prediction and summary generation tasks, we performed a comprehensive ablation study, with results detailed in Table \ref{tab:curriculum sft}. Our analysis validates a two-part hypothesis regarding the roles of the different stages. First, atomic tag prediction training (Stage 1 and Stage 2) is indispensable to build foundational atomic reasoning capabilities. The model trained on Stage 1 data (Row 2) achieves 39.08\% Pass@1 improvements (0.5030->0.6996) on  HPR-Bench\textsubscript{tag} compared to the backbone (Row 1). Subsequently, integrating Stage 2 data into the training process further yields a significant boost in atomic tag prediction performance (e.g., Pass@1 of 0.7326). This demonstrates that these initial stages are essential to equip the model with the core ability to infer individual user attributes.

Second, the key contribution of the Stage 3 summary generation training is not to further refine atomic accuracy, but to teach the model how to cohesively organize and present these attributes. This is evidenced by two key observations. On one hand, adding the summary stage (Row 4 vs. Row 3) yields no consistent improvement in atomic tag accuracy, with Pass@1 increasing while Pass@5 and Pass@10 decreasing on HPR-Bench\textsubscript{tag}. On the other hand, it dramatically improves the quality of the final composite profile summary. On HPR-Bench\textsubscript{sum}, the $\text{Acc}_\text{Ex}$ score surges from 0.4926 to 0.6595, and the $\text{COV}_\text{Ex}$ score increases significantly from 0.8206 to a remarkably high 0.9563.

Finally, multi-stage curriculum-driven SFT strategy further enhances the final composite profile summary performance. We compare naive mixed-data training (Row 4) and multi-stage curriculum-driven training (Row 5), ensuring an equal amount of data in each variant. By organizing the same data into a curriculum (training on Stage 1 first, then Stage 2 and finally Stage 3), the multi-stage curriculum-driven model reaches a highest $\text{Acc}_\text{Ex}$ of 0.6899 (+4.6\%), demonstrating that a progressive, easy-to-hard learning sequence is the most effective strategy for enhancing the model's final composite profile summary capabilities. More crucially, it still maintains robust performance on atomic tag prediction, which achieves highest Pass@5 (0.8911) and Pass@10 (0.9230) on tag prediction, indicating a substantial increase in the model's theoretical upper bound. This endows the model with a vast problem-solving space and a rich repertoire of effective reasoning paths, providing an ideal foundation for reinforcement learning (RL).

In conclusion, our multi-stage curriculum operates synergistically: Stage 1 and 2 build the "what" (the ability to predict facts), while Stage 3 teaches the "how" (the ability to structure these facts into a complete and coherent narrative).

\subsubsection{Synergistic Effect of Sample and Group Filtering.}
\begin{table}[htbp]
\footnotesize
        \centering
        \caption{Ablation Study of RL on $\text{HPR-Bench}_\text{sum}$.}
        \label{tab:ab_RL_sum}
        \begin{tabular}{p{16em}cccccc}
        \toprule
        \multirow{2}[4]{*}{\textbf{Models}} & \multicolumn{6}{c}
        {\textbf{Quantitative evaluation}} \\
        \cmidrule{2-7}    \multicolumn{1}{c}{} & \multicolumn{1}{p{3em}}{$\textbf{Acc}_\textbf{Ex}$} & \multicolumn{1}{p{3em}}{$\textbf{COV}_\textbf{Ex}$} & \multicolumn{1}{p{3em}}{\textbf{BLEU\_2}} & \multicolumn{1}{p{3em}}{\textbf{BLEU\_4}} & \multicolumn{1}{p{4em}}{$\textbf{Score}_\textbf{sim}$} \\
        \midrule
        SFT & 	0.6899 &  	0.9587  & 	0.1631  & 	0.0378 &  	  	0.7854 \\
        GRPO & 	0.7046 &  	0.9371  & 	0.1737  & 	0.0621 &  	  	0.8129 \\
        GRPO with sample-level & 	0.6949 &  	0.9362  & 	0.1833  & 	0.0604 &  	  	0.7896 \\
        GRPO with group-level & 	0.7323 &  	0.9681  & 	\textcolor[rgb]{ 1,  0,  0}{0.2038}  & 	0.0649 &  	 	0.8089 \\
        DF-GRPO & \textcolor[rgb]{ 1,  0,  0}{0.7528} & \textcolor[rgb]{ 1,  0,  0}{0.9747} & 0.1931 & \textcolor[rgb]{ 1,  0,  0}{0.0691} & \textcolor[rgb]{ 1,  0,  0}{0.8215} \\
        \hline
        \end{tabular}
\end{table}

To evaluate the effectiveness of our proposed DF-GRPO strategy, we conducted an ablation study focusing on the sample-level and group-level filtering mechanisms during the reinforcement learning phase. The results, summarized in Table \ref{tab:ab_RL_sum}, reveal several insights into how these filtering layers contribute to performance on $\text{HPR-Bench}_\text{sum}$.

Initially, standard GRPO shows a noticeable improvement over the SFT baseline, particularly regarding extraction accuracy $\text{Acc}_\text{Ex}$ (increasing from 0.6899 to 0.7046) and $\text{Score}_\text{sim}$ (increasing from 0.7854 to 0.8129). The introduction of our dual-filtering mechanism further extends these performance boundaries.

The importance of \textbf{Group-level Filtering} is highlighted by the performance degradation observed when it is removed (GRPO with sample-level). In this configuration, the extraction accuracy $\text{Acc}_\text{Ex}$ falls to 0.6949, which is only marginally better than the SFT baseline. This suggests that filtering out low-quality groups is essential for stable training. Specifically, in cases where the model is fundamentally confused or reward signals are indistinguishable, the lack of filtering introduces significant noise into the optimization process, ultimately leading to suboptimal convergence.

Furthermore, \textbf{Sample-level Filtering} plays a crucial role in refining the advantage calculation within each group. While the ``DF-GRPO with sample-level'' variant performs reasonably well in terms of BLEU scores, the full \textbf{DF-GRPO} achieves the highest $\text{Acc}_\text{Ex}$ (0.7528) and $\text{COV}_\text{Ex}$ (0.9747). This indicates that by removing noisy trajectories within a group before computing the mean reward, we provide a more accurate and stable advantage signal for the policy update.

Ultimately, the synergy between these two levels of filtering enables DF-GRPO to achieve the best overall performance across nearly all metrics. By simultaneously discarding low-fidelity groups and pruning outlier samples, our method ensures that the model learns from the most informative and high-fidelity reasoning paths available.

\section{Discussion: Incremental Profiling}
While our experiments demonstrate that UserGPT can effectively compress long-term user histories into a concise summary of just a few hundred tokens while preserving critical personalization information, reaching the compression ratio of 92\%, two major challenges arise when considering large-scale industrial deployment.
\vspace{-0.5cm}
\begin{itemize}
    \item \textbf{Efficiency Bottleneck}: Re-processing years of history to incorporate a single day of new activity results in over 99\% redundant computation. This is computationally prohibitive for hundreds of millions of users.
    \item \textbf{Temporal Conflict}: Static inference struggles to balance stable, long-term attributes (e.g., gender) with dynamic, short-term interests (e.g., current intent). Historical data can often "dilute" and obscure recent, high-signal behavioral shifts.
\end{itemize}

\begin{figure}[htbp]
    \centering
    \includegraphics[width=0.95\textwidth]{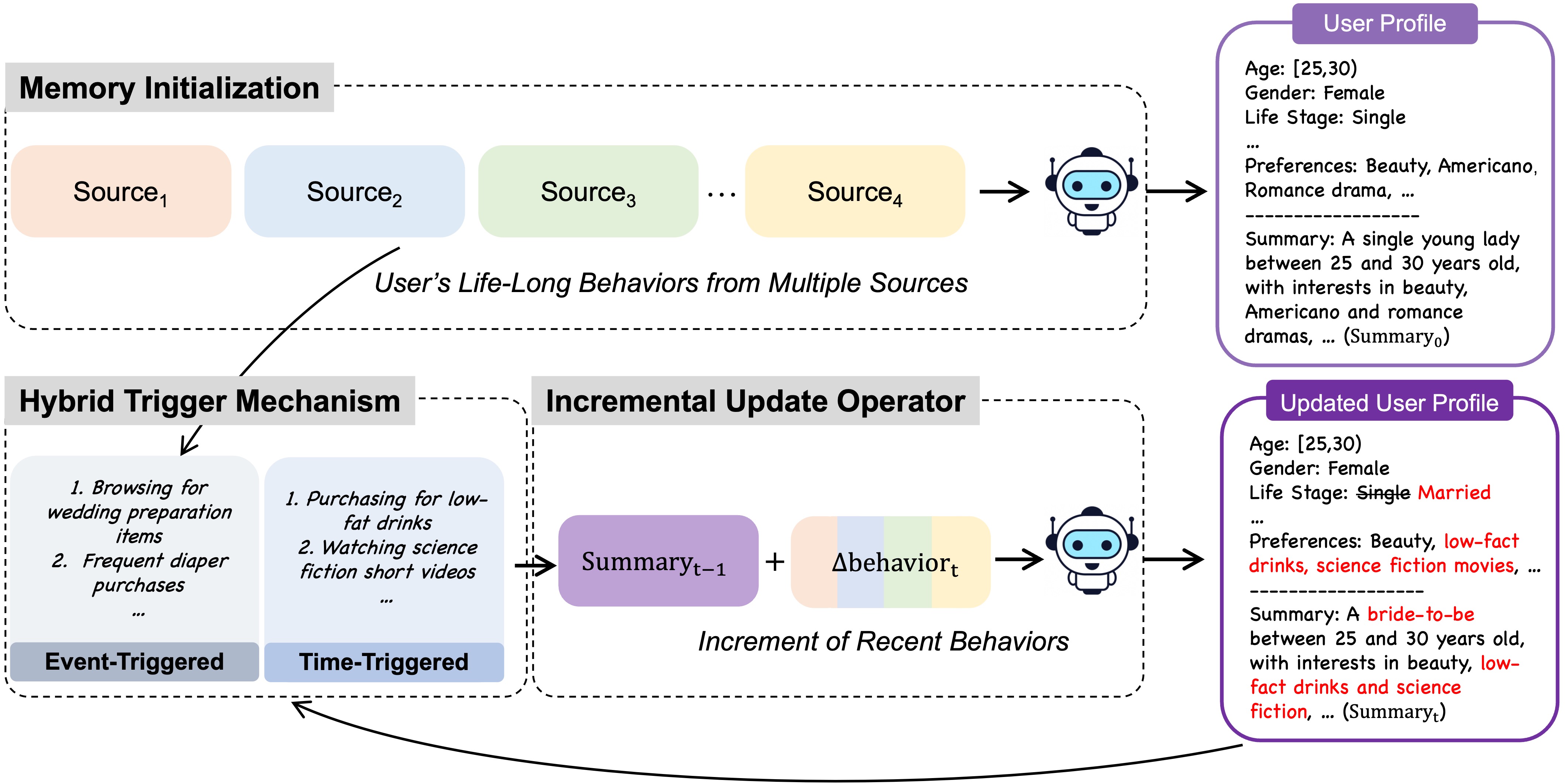}
    \caption{The proposed \textbf{Incremental Profiling} paradigm, including three components: initialization, update and triggering.}
    \label{fig: system pipeline}
\end{figure}
\vspace{-0.5cm}

To address these challenges, we envision a future direction centered on an Incremental Profiling paradigm, as illustrated in Figure  \ref{fig: system pipeline}. This approach treats the generated summary as a dynamic, pluggable memory for AI agents. The process would begin with a \textbf{profile initialization step,} where UserGPT performs a one-time, offline compression of a user's complete historical logs into a foundational summary$Summary_0$. Subsequently, instead of full re-computation, the profile would be updated recursively. This \textbf{incremental update} would involve UserGPT assimilating recent behavioral changes($\Delta Behavior_t$) into the previous summary($Summary_{t-1}$), as formulated in Eq. \ref{eq:incremental mechanism}, effectively performing an 'Assimilate-Reason-Rectify' cycle to reflect both stable traits and dynamic persona shifts. To balance cost and timeliness, these updates would be governed by a hybrid trigger mechanism:
\vspace{-0.5cm}
\begin{itemize}
    \item \textbf{Event-Triggered}: Instant updates for high-value signals, such as changes in shipping address, ensuring immediate response to life-stage shifts.
    \item \textbf{Time-Triggered}: Periodic updates (e.g., weekly) to capture gradual preference drifts and ensure the profile remains synchronized with recent activities.
\end{itemize}
\begin{equation}
    Summary_t=UserGPT(Summary_{t-1}+\Delta Behavior_t)
    \label{eq:incremental mechanism}
\end{equation}

This evolving, natural language summary is natively suited to empower next-generation AI agents. Upon receiving explicit user authorization, it could instantly eliminate the "cold start" problem for new applications; for example, a travel agent app could immediately know a user's preference for ``non-smoking hotels with a gym''. More profoundly, it enhances an agent's ability to understand latent intent and tailor its communication strategy. When asked to ``explain a black hole'', the agent could provide a technically dense answer to a user identified as a physicist, while offering a simple analogy like a ``cosmic drain'' to a user whose summary indicates an interest in popular science. This transforms the user profile from a static set of tags into an evolving cognitive record, serving as a foundational memory layer for truly personalized AI interactions.

\section{Conclusion \& Future Work}
\label{sec:conclusion and future work}
In this paper, we presented \textbf{UserGPT}, a principled framework that redefines LLM-based user profiling, spanning the entire lifecycle from data preparation to production deployment. We begin by confronting the dual obstacles of data scarcity and usability, proposing a novel data curation pipeline that combines a \textbf{User Behavior Simulation Engine} with \textbf{Data-Centric Semantization} to create a robust and LLM-friendly data foundation. Upon this foundation, we developed a \textbf{curriculum-driven post-training paradigm} for UserGPT to systematically enhance its cognitive abilities, tackling the inherent limitations of LLMs in temporal reasoning and logical consistency. The effectiveness of our approach was rigorously validated on our newly constructed \textbf{HPR-Bench}, a comprehensive benchmark for complex persona reasoning. Taken together, this work establishes a complete, end-to-end methodology for moving beyond the fragmented, tag-based profiles of the past toward a more holistic, narrative-driven understanding of users.

While this holistic framework establishes a viable and effective path toward cognitively rich user understanding, significant advances remain required on the journey toward truly human-like comprehension. Our exploration illuminates several inherent challenges and, in doing so, charts a clear course for future inquiry. These frontiers represent not just incremental improvements, but fundamental steps toward the next generation of intelligent systems.

First, a key challenge remains in \textbf{enhancing implicit and temporal reasoning}. While our model shows strong performance, future work should focus on enabling more sophisticated, multi-hop logical inference from sparse historical events. For instance, this would involve training models to infer dynamic attributes, such as a child’s current age, by reasoning over fine-grained details from past behaviors (e.g., shoe sizes purchased years ago).

Second, and relatedly, is the need for \textbf{advanced hallucination suppression}. Current models can occasionally over-interpret sparse or noisy user signals, leading to profile inconsistencies. A critical future direction is to develop mechanisms that guide the model to ground its inferences in persistent, life-stage-consistent evidence within a user's history, rather than on sporadic, outlier behaviors. This would significantly improve the model's factual faithfulness.

Furthermore, as LLMs' context windows expand, future work must address the challenge of \textbf{effectively utilizing longer contexts}. The goal extends beyond simply accommodating more data; it involves developing strategies to help models retain and focus on critical fine-grained details amidst vast amounts of information, mitigating issues like information overload and the "lost-in-the-middle" problem.

Finally, a significant frontier is \textbf{multimodal fusion for user profiling}. Extending our text-based framework to incorporate visual signals, such as image and video browsing history, would allow the model to capture users' aesthetic tastes and visual preferences. This would pave the way for a truly holistic and nuanced understanding of the user.

In conclusion, our contributions not only deliver a state-of-the-art solution but also offer a new paradigm for the field. By addressing the core challenges across the data, modeling, and system layers, and by outlining these critical directions for future research, we pave the way for a new generation of truly personalized, intelligent, and context-aware user-agent interactions.

\addcontentsline{toc}{section}{References}
\bibliographystyle{myabbrvnat}
\nobibliography*
\bibliography{reference}

\clearpage

\appendix
\section*{Appendix}
\section{User Profile Prompt Template}
\label{appendix:prompt}

As discussed in Section \ref{subsubsection:Prompt Engineering}, the following template illustrates the actual prompt utilized in our experiments. While the original experiments were conducted in Chinese to align with the dataset's primary language, an English translation is provided here for accessibility.

\begin{tcolorbox}[
    enhanced,
    breakable,     
    colback=prompt_bg, 
    colframe=prompt_frame,
    fonttitle=\bfseries\large\sffamily,
    title={Example of Full Prompt},
    attach boxed title to top center={yshift=-2mm},
    boxed title style={
        colback=prompt_frame,
        colframe=prompt_frame,
        arc=4pt,
    },
    arc=4pt,
    boxrule=1pt,
    left=10pt, right=10pt, top=10pt, bottom=10pt,
    pad at break=0pt, 
    before upper={\setlength{\parindent}{0pt}},
]
\small
\# Role \\
You are a professional user profile analyst. Given a user's behavioral data, provide accurate profile attributes with concise reasoning steps.

\vspace{5pt}
\# Task \\
Objective: Please construct a comprehensive and accurate user profile based on the provided anonymized behavioral data. Follow these steps:
\begin{enumerate}[leftmargin=20pt, nosep]
    \item Reason step-by-step based on the provided behavior samples.
    \item Format the output.
\end{enumerate}

Constraint:

\begin{itemize}[leftmargin=15pt, itemsep=4pt, parsep=0pt]
    \item Evidence Threshold: Attribute inference must be supported by multiple independent behavioral signals. Single interactions are insufficient for any conclusion. 
    \item Ambiguous Roles: Distinguish between students, faculty, and merchants by examining age and consumption patterns.
    \item Product Versatility: Acknowledge that items like ``baby wipes'' or ``non-perforated hooks'' are versatile. Single interactions are insufficient for profiling, multiple independent clues are required to ensure logical rigor.
    \item Conservatism Principle: When evidence is ambiguous, default to the more general category or mark the attribute as "Unknown" rather than over-infer. 
    \item Consistency Check: Ensure logical consistency across the profile (e.g., active students are rarely employed full-time, parents are generally married).
    \item Intent Analysis: Distinguish between inherent behavioral signals and marketing-driven noise. Search and carting behavior indicate demand, even if the final transaction occurred offline.
    \item ...\\
    (\textit{The full prompt template containing domain-specific inference heuristics is proprietary and omitted from this report for confidentiality.})
\end{itemize}

\vspace{5pt}
\# Input \\
Behavioral Data:{\{Insert multi-source user behavior data\}} 

\# Output \\
Format:Return a valid JSON object as follows:
\begin{lstlisting}[basicstyle=\ttfamily\scriptsize, numbers=none]
{
  "I. Demographic and Social Attributes": {
    "Gender": "",
    "Age group": "",
    "Education": "",
    "Is currently a student": "",
    "Life stage": "",
    "Household Context": "",
    "Occupation": "",
    ...
  },
  "II. Lifestyle Tier": "",
  "III. Geographic Context": "",
  "IV. Consumption and Lifestyle": {
    "Hobbies": [
      "List up to 5 high-confidence, long-term, and stable hobbies in string list format, if none, do not output"
    ],
    "Brand Preference": "",
    "Pet": {
      "Whether to keep pets": "",
      "Pet type": [
        "List high-confidence pet types in the form of a string list. If none, do not output"
      ]
    },
    ...
  },
  "V. Composite User Profile Summary": "Based on the aforementioned dimensions , provide a 300-500 word summary of user profile."
}
\end{lstlisting}
\end{tcolorbox}

\section{Case Study}
\label{appendix:case study}
We show an example to generate user profile summary with UserGPT. We construct the complete input instruction by filling the corresponding placeholders ''{Insert multi-source user behavior data}'' in the full prompt template with the concrete values of user behavior data. The output response includes reasoning and final answer. In the following, we display the detailed User behavior data and response of UserGPT. (\textbf{Note: The presented case is synthesized by the User Behavior Simulation Engine mentioned in Section \ref{sec:behavior simulation} and does not correspond to any specific real individual.})

\begin{evolvingbox}[title={User behavior data}]
\textbf{Current time}: 2026-03-01 \\
\noindent
\textbf{E-commerce}:
\noindent
\begin{itemize}
\item  2018 Purchase | Diapers for Baby Girls,  ...
\item 2019 Purchase | Boys' Cotton Jacket, Smart Rice Cooker, 
\item 2020 Search/Click | Schoolbag for Primary School Children, Minimalist Shoe Cabinet, ...
\item 2025-03 Added Cart | Tail Light for Benz, Junior High School Mathematics Textbook, ...
\item 2026-02-17 Purchase | Dog Food for Puppy, ...
\item ...
\end{itemize}
\noindent
\textbf{OTA}:
\noindent
\begin{itemize}

\item ...
ng the Family Fortune, Ex-Wife Begs for Forgiveness)
\end{itemize}
\textbf{Food Delivery}:
\noindent
\begin{itemize}
\item ...
\end{itemize}
\end{evolvingbox}

\begin{evolvingbox}[title={Response of UserGPT}]
\textbf{\#Reasoning:} \\

\textbf{1.Life Stage Analysis}\\
\textcolor{blue}{\uline{Baby \& children's products appear frequently in e-commerce purchase records}},  including ``Diapers for Baby Girls'', ..., confirmed as \textcolor{red}{"Family-oriented"}.\\
...\\
(\textit{Detailed reasoning chains are omitted for brevity. The focus here is on demonstrating the structured inference workflow rather than specific attribute values.})

\textbf{\#Answer:}
\begin{lstlisting}[basicstyle=\ttfamily\tiny, numbers=none]
{
  "I. Demographics and Social Attributes": {
    "Age Group": "40-44",
    "Gender": "Female",
    "Is currently a student": "No",
    "Life stage": "Family-oriented",
    "Household Context": "Multi-member household with a daughter at Primary School",
    "Occupation": ...,
  },
  "II. Lifestyle Tier": "Quality-conscious consumer",
  "III. Geographic Context": ...,
  "IV. Consumption and Lifestyle": {
    "Hobbies": [
      "Parenting",
      "Education",
      "Family Life"
    ],
    "Pet": {
      "Whether to keep pets": "Yes",
      "Pet types": [
        "Dog"
      ]
    }
  },
  "V. Composite User Profile Summary": "The user is a female at the age of 40-44, ..."
}
\end{lstlisting}
\end{evolvingbox}

\section{Additional Evaluation Results}
\label{appendix:Additional Evaluation Results}
\begin{table}[H]
  \tiny
  \setlength{\tabcolsep}{0.8pt}  
  \centering
  \caption{Detailed Pass@1,Pass@5 and Pass@10 on $\text{HPR-Bench}_\text{tag}$. Within each column, red represents the highest value, followed by blue. (LS: Life Stage, HC: Household Context, LI: Lifestyle Indicators, EP:Educational \& Professional Background, GC: Geographic Context) }
    \begin{tabular}{p{6em}rrrrrrrrrrrrrrrrrrrrrr}
    \toprule
    \multirow{2}[4]{*}{\textbf{Model}} & \multicolumn{1}{p{3.5em}}{\multirow{2}[4]{*}{\textbf{Pass@1}}} & \multicolumn{1}{p{3.5em}}{\multirow{2}[4]{*}{\textbf{Pass@5}}} & \multicolumn{1}{p{3.5em}}{\multirow{2}[4]{*}{\textbf{Pass@10}}} & \multicolumn{3}{c}{\textbf{LS}} &       & \multicolumn{3}{c}{\textbf{HC}} &       & \multicolumn{3}{c}{\textbf{LI}} &       & \multicolumn{3}{c}{\textbf{EP}} &       & \multicolumn{3}{c}{\textbf{GC}} \\
\cmidrule{5-7}\cmidrule{9-11}\cmidrule{13-15}\cmidrule{17-19}\cmidrule{21-23}    \multicolumn{1}{l}{} & \multicolumn{1}{p{3.5em}}{} & \multicolumn{1}{p{3.5em}}{} & \multicolumn{1}{p{3.5em}}{} & \multicolumn{1}{c}{\textbf{pass@1}} & \multicolumn{1}{c}{\textbf{pass@5}} & \multicolumn{1}{c}{\textbf{pass@10}} &       & \multicolumn{1}{c}{\textbf{pass@1}} & \multicolumn{1}{c}{\textbf{pass@5}} & \multicolumn{1}{c}{\textbf{pass@10}} &       & \multicolumn{1}{c}{\textbf{pass@1}} & \multicolumn{1}{c}{\textbf{pass@5}} & \multicolumn{1}{c}{\textbf{pass@10}} &       & \multicolumn{1}{c}{\textbf{pass@1}} & \multicolumn{1}{c}{\textbf{pass@5}} & \multicolumn{1}{c}{\textbf{pass@10}} &       & \multicolumn{1}{c}{\textbf{pass@1}} & \multicolumn{1}{c}{\textbf{pass@5}} & \multicolumn{1}{c}{\textbf{pass@10}} \\
    \midrule
    \rowcolor{gray!10} \multicolumn{23}{c}{Non-Thinking Models} \\
    \midrule
    Qwen3-8B Non-Thinking & 0.4612  & 0.5071  & 0.5154  & 0.5807  & 0.6143  & 0.6193  &       & 0.3281  & 0.3796  & 0.3899  &       & 0.6091  & 0.6497  & 0.6546  &       & 0.4282  & 0.4514  & 0.4533  &       & 0.4130  & 0.4707  & 0.4827  \\
    Qwen3-235B-A22B-Instruct-2507 & 0.5244  & 0.5912  & 0.6101  & 0.5944  & 0.6371  & 0.6541  &       & 0.4140  & 0.5079  & 0.5346  &       & 0.5194  & 0.5377  & 0.5420  &       & 0.5584  & 0.5898  & 0.5971  &       & 0.5341  & 0.6306  & 0.6564  \\

    \midrule
    \rowcolor{gray!10} \multicolumn{23}{c}{Thinking Models} \\
    \midrule
    Qwen3-8B Thinking & 0.5035  & 0.6881  & 0.7385  & 0.6224  & 0.8062  & 0.8505  &       & 0.3831  & 0.5892  & 0.6494  &       & 0.6244  & 0.7641  & 0.7989  &       & 0.4770  & 0.5894  & 0.6259  &       & 0.4599  & 0.6757  & 0.7349  \\
    Qwen3-235B-A22B-Thinking-2507 & 0.6434  & 0.7854  & 0.8221  & 0.6411  & 0.7990  & 0.8369  &       & 0.4135  & 0.5718  & 0.6247  &       & 0.6438  & 0.7750  & 0.8062  &       & 0.6261  & 0.8046  & 0.8612  &       & 0.7478  & 0.8702  & 0.8960  \\
    Qwen3.5-397B-A17B & 0.7165  & 0.8311  & 0.8614  & 0.7688  & 0.8905  & 0.9138  &       & 0.5466  & 0.7135  & 0.7636  &       & 0.6912  & 0.8042  & 0.8315  &       & 0.7071  & 0.8270  & 0.8658  &       & 0.7804  & 0.8688  & 0.8923  \\
    Qwen3.5-plus & 0.7154  & 0.8333  & 0.8661  & 0.7744  & 0.9023  & 0.9277  &       & 0.5400  & 0.7158  & 0.7746  &       & 0.6923  & 0.7954  & 0.8173  &       & 0.7047  & 0.8318  & 0.8737  &       & 0.7781  & 0.8706  & 0.8977  \\
    Kimi-K2-Thinking & 0.6556  & 0.8548  & 0.8956  & 0.7083  & \textcolor[rgb]{ 1,  0,  0}{0.9191 } & \textcolor[rgb]{ 1,  0,  0}{0.9544 } &       & 0.5380  & 0.7784  & 0.8343  &       & 0.6455  & 0.8229  & 0.8535  &       & 0.6350  & 0.8225  & 0.8596  &       & 0.6937  & 0.8829  & 0.9249  \\
    Kimi-K2.5 & 0.7064  & 0.8421  & 0.8758  & 0.7467  & 0.8657  & 0.8936  &       & 0.5380  & 0.7730  & 0.8220  &       & 0.6866  & 0.7917  & 0.8166  &       & 0.6590  & 0.7701  & 0.8073  &       & 0.7834  & 0.9038  & 0.9361  \\
    DeepSeek-R1-0528 & 0.6613  & 0.8012  & 0.8380  & 0.6412  & 0.7827  & 0.8164  &       & 0.4752  & 0.6621  & 0.7253  &       & 0.6664  & 0.7907  & 0.8226  &       & 0.6786  & 0.8318  & 0.8702  &       & 0.7424  & 0.8646  & 0.8928  \\
    DeepSeek-V3.2 & 0.6684  & 0.8071  & 0.8439  & 0.6842  & 0.8486  & 0.8819  &       & 0.4981  & 0.6716  & 0.7225  &       & 0.6753  & 0.7959  & 0.8256  &       & 0.6693  & 0.8140  & 0.8528  &       & 0.7314  & 0.8503  & 0.8850  \\
    GLM-4.7 & 0.6938  & 0.8070  & 0.8386  & 0.7736  & 0.8719  & 0.8982  &       & 0.5200  & 0.6729  & 0.7245  &       & \textcolor[rgb]{ .267,  .447,  .769}{0.7016 } & 0.7666  & 0.7892  &       & 0.6465  & 0.7629  & 0.7924  &       & 0.7443  & 0.8666  & 0.8964  \\
    GLM-5 & 0.7186  & 0.8467  & 0.8768  & 0.7691  & 0.8985  & 0.9247  &       & 0.5345  & 0.7138  & 0.7624  &       & \textcolor[rgb]{ 1,  0,  0}{0.7098 } & 0.8144  & 0.8399  &       & 0.7037  & 0.8371  & 0.8761  &       & \textcolor[rgb]{ 0,  .439,  .753}{0.7838 } & 0.8979  & 0.9214  \\
    Qwen3.6-Plus & \textcolor[rgb]{ 1,  0,  0}{0.7329 } & 0.8517  & 0.8781  & \textcolor[rgb]{ 1,  0,  0}{0.7907 } & \textcolor[rgb]{ .267,  .447,  .769}{0.9049 } & \textcolor[rgb]{ .267,  .447,  .769}{0.9265 } &       & 0.5965  & 0.7982  & 0.8479  &       & 0.6991  & 0.7950  & 0.8160  &       & 0.6694  & 0.7999  & 0.8370  &       & \textcolor[rgb]{ 1,  0,  0}{0.7993 } & 0.8908  & 0.9088  \\

    \midrule
    $\textbf{UserGPT}_\textbf{SFT}$ & \textcolor[rgb]{ 0,  .439,  .753}{0.7325 } & \textcolor[rgb]{ 1,  0,  0}{0.8911 } & \textcolor[rgb]{ 1,  0,  0}{0.9230 } & 0.7776  & 0.8838  & 0.9059  &       & \textcolor[rgb]{ 1,  0,  0}{0.6088 } & \textcolor[rgb]{ 1,  0,  0}{0.8365 } & \textcolor[rgb]{ 1,  0,  0}{0.8837 } &       & 0.6999  & \textcolor[rgb]{ 1,  0,  0}{0.8377 } & \textcolor[rgb]{ 1,  0,  0}{0.8808 } &       & \textcolor[rgb]{ 1,  0,  0}{0.8211 } & \textcolor[rgb]{ 1,  0,  0}{0.8996 } & \textcolor[rgb]{ 1,  0,  0}{0.9203 } &       & 0.7548  & \textcolor[rgb]{ 1,  0,  0}{0.9380 } & \textcolor[rgb]{ 1,  0,  0}{0.9662 } \\
    UserGPT & 0.7306  & \textcolor[rgb]{ .267,  .447,  .769}{0.8766 } & \textcolor[rgb]{ .267,  .447,  .769}{0.9062 } & \textcolor[rgb]{ 0,  .439,  .753}{0.7879 } & 0.8834  & 0.9059  &       & \textcolor[rgb]{ .267,  .447,  .769}{0.5962 } & \textcolor[rgb]{ .267,  .447,  .769}{0.7937 } & \textcolor[rgb]{ .267,  .447,  .769}{0.8388 } &       & 0.6945  & \textcolor[rgb]{ .267,  .447,  .769}{0.8315 } & \textcolor[rgb]{ .267,  .447,  .769}{0.8672 } &       & \textcolor[rgb]{ .267,  .447,  .769}{0.8164 } & \textcolor[rgb]{ .267,  .447,  .769}{0.8934 } & \textcolor[rgb]{ .267,  .447,  .769}{0.9071 } &       & 0.7546  & \textcolor[rgb]{ .267,  .447,  .769}{0.9238 } & \textcolor[rgb]{ .267,  .447,  .769}{0.9516 } \\
    \bottomrule
    \end{tabular}%
  \label{tab:addlabel}%
\end{table}%

\end{CJK*}
\end{document}